# Experimental Evidence for Defect Tolerance in Pb-Halide Perovskite

Naga Prathibha Jasti[1,2], Igal Levine[3], Yishay (Isai) Feldman[2], Gary Hodes[2], Sigalit Aharon[2*] and David Cahen[1,2*]

[1]Bar Ilan University, Ramat Gan, 5290002, Israel
[2]Weizmann Institute of Science, Rehovot, 7610001, Israel
[3]Division Solar Energy, Helmholtz-Zentrum Berlin für Materialien und Energie GmbH, Berlin, 12489, Germany

*david.cahen@weizmann.ac.il, sigalit.aharon@weizmann.ac.il

**Abstract**

The term defect tolerance (DT) is used often to rationalize the exceptional optoelectronic properties of Halide Perovskites (HaPs) and their devices. Even though DT lacked direct experimental evidence, it became a "fact" in the field. DT in semiconductors implies that structural defects do not translate to electrical and optical effects (e.g., due to charge trapping), associated with such defects. We present the first direct experimental evidence for DT in Pb-HaPs by comparing the structural quality of 2-dimensional (2D), 2D-3D, and 3D Pb-iodide HaP crystals with their optoelectronic characteristics using high-sensitivity methods. Importantly, we get information from the materials' bulk, because we sample at least a few hundred nanometers, up to several micrometers, from the sample's surface, which allows for assessing intrinsic bulk (and not only surface-) properties of HaPs. The results point to DT in 3D, 2D-3D, and 2D Pb-HaPs. Overall, our data provide an experimental basis to rationalize DT in Pb-HaPs. These experiments and findings can guide the search for, and design of other materials with DT.

**Significance**
For progress in science, distinguishing between ideas or suggestions and solid experimental facts is critical, especially if it concerns a key issue. Such is the case for the widespread use of "defect tolerance" (DT) to explain the exceptionally good properties of Halide Perovskites (HaPs) for solar cells, LEDs and detectors. Although often invoked for HaPs, DT lacked till now direct experimental evidence (and had mixed support from computational theory). We find the missing experimental evidence for DT by comparing the structural quality, as a measure of structural defects, with its optoelectronic properties, as a measure of optoelectronic (active) defects, that normally result from structural defects. Our direct evidence for DT can guide the search for, and design of new materials with DT.

**Main text**

**Introduction**

Halide perovskite (HaP) semiconductors hold significant potential for applications in several optoelectronic device types due to their excellent relevant properties, which express fascinating fundamental materials chemistry and physics (1–6). Remarkably, also HaP *polycrystalline* layers, even if deposited at near-room temperature and ambient pressure from solution or by printing, can show such excellent optoelectronic material properties and yield highly efficient devices (7, 8). This behaviour is notable because film preparation under these conditions should result in high structural, crystallographic defect concentrations, which appears incompatible with the high level of crystallographic order that such films can show. Structural defects are distortions or dislocations of atoms (ions) from the unit cell of the theoretical crystal structure, and include 0D (point), 1D (line), 2D (planar) and 3D (bulk) defects. This result is surprising as it is quite different from what is the case with "classical semiconductors", (9–12) where surfaces and grain boundaries need to be carefully passivated and structural defects eliminated or mitigated in post-synthesis steps, to reach a similar performance. Such a combination of preparation/properties/performance combination is unheard of for today's commercially used, as well as non-commercial emergent inorganic semiconductors, and till now, seems unique for HaPs. Thus, it has created enormous interest in understanding what causes this combination (13–16).

Commonly this unique feature of HaPs has been explained by defect-tolerance (DT), (17–22) based on theoretical models and computations, i.e., the materials were "declared" to be "defect tolerant" semiconductors (10) (see next paragraph). A recent high-level computational electronic structure study concluded, though, against DT (23). Strong computational theory support for DT came from combining molecular dynamics with DFT computations, showing large (up to 1 eV) picoseconds fluctuations in in-gap defect levels (24, 25), which facilitates a DT mechanism. Another rationalization is the coupling of the electronic carriers to low-frequency phonon modes (resulting in small bulk moduli and therefore a soft lattice), making non-radiative recombination a multi-phonon process (26–28). While experimental evidence for DT remained indirect or absent (specifically as an intrinsic property of the bulk material) (29), such evidence for self-healing (SH) of defects in HaP materials (single crystals and poly-crystalline films) has been accumulating, from our group and others (30–36). Recently we compared the two explanations (DT vs. SH) for the exceptional qualities of HaPs and concluded that DT alone cannot explain HaP behavior (37). We noted that especially transient DT, acting in conjunction with SH, cannot be excluded, and, as a caveat, that, as yet clear experimental results for DT (beyond "consistent with") remain lacking.

As pointed out, e.g., in ref. 37 (37), experimentally showing defect tolerance is challenging, in part because the term 'defect tolerance', as used in the context of the optoelectronic properties of HaPs, is somewhat ambiguous. In a theoretical work on chalcopyrites (CIGS), the "electrically benign nature of structural defect pairs" that those materials can sustain due to "structural tolerance to large off-stoichiometry" was noted (11). For HaPs, DT has been justified by their high dielectric constant ($\sim$ 3) times larger than for CIGS or CdTe thin films), because it reflects their ability to effectively screen electrostatic perturbations. Defect-defect interactions were central in those works, as they were thought to change in-gap defect levels to resonances in the bands (38). There are other definitions for the term 'DT', but the most operational one (meaning, one that translates directly to device performance) is the following: Semiconductors such as HaPs (especially Pb-HaPs), are 'defect tolerant' if they show good optoelectronic quality, even if they (are expected to) have a high structural defect density (including 0D, 1D, 2D, and/or 3D structural defects) (10).

The rationale for such a definition is that high optoelectronic quality normally reflects a low density of intrinsic structural defects because such defects can act as efficient electronic charge traps or recombination centers. Notably, the direct implication of this operative definition is that one must provide both structural evidence (from evaluating the structural quality of the material, bulk, or bulk and surfaces) and optoelectronic evidence (assessed via a material's properties, such as optical absorption or emission, electronic transport, etc.) to test proposed defect tolerance of HaPs. In fact, this is correct also when referring to other definitions of DT, as the term relates the effects of existing intrinsic structural defects to the optoelectronic characteristics of HaPs (38–40).

To yield direct experimental evidence for DT in Pb-HaPs, we report and compare the structural quality of pure 3D, 2D, and 2D-3D Pb-HaP single crystals to optoelectronic characteristics that are sensitive to structural defects. We compare a family of Pb-iodide HaPs:

- pure 2D HaP - $(C_4H_9NH_3)_2PbI_4$ or $BA_2PbI_4$ (C4N1),
- 2D-3D HaPs - $(C_4H_9NH_3)_2(CH_3NH_3)Pb_2I_7$ or $BA_2MAPb_2I_7$ (C4N2)
  and $(C_4H_9NH_3)_2(CH_3NH_3)_2Pb_3I_{10}$ or $BA_2MA_2Pb_3I_{10}$ (C4N3),
and the widely studied, archetypical
- 3D HaP - $(CH_3NH_3)PbI_3$ or $MAPbI_3$ (MAPI).

While most of the HaP-based potential applications use polycrystalline thin films, fundamental understanding requires at least to start with more controllable samples, wherever possible, i.e., high-quality single crystals or epitaxial films. As the latter is not (yet) readily available in macroscopic sizes, if at all, we, as others, have turned to single crystals for fundamental studies. Thus, single crystals of Pb-HaPs with different dimensionalities were grown and are analyzed here, to try to assess possible differences between them in terms of the relationship between structural and optoelectronically active imperfections, if any (also building on earlier studies of ours concerning self-healing differences between 2D and 3D HaPs (33)).

# Main text

To assess the quality of such single crystals, one normally relies heavily on X-ray diffraction (XRD) measurements with the θ-2θ Bragg-Brentano (θ-2θ, for short) geometry, by identifying the presence of a single set of peaks, corresponding to a single phase with one clear crystallographic orientation. While certainly a necessary condition, it is far from being a sufficient one, as the common θ-2θ XRD experiments are insufficiently sensitive to slight structural inhomogeneities. We evaluate the single crystal quality of these materials using additional XRD measurements, apart from the qualitative results from θ-2θ geometry, namely quantitative rocking curves (RC, see Figure S1, SI). The RC experiments include repetitions at different φ values (see Figure S1, SI). These extra measurements inform on possible tilts of crystallographic planes with respect to the single crystal surface (out-of-plane rotations), as well as on twists about the substrate normal (in-plane rotations). In addition, the tails of the RC also reflect point defects at > $10^{13}$/cm$^3$ densities (estimate for single crystal Si; cf. ref. 41 (41)), i.e., 1-3 orders of magnitude above the doping /trap densities that have been deduced for high quality HaP single crystals. Overall, the RC width, specifically its full width at half maximum (FWHM), as well as any differences between the RC shapes at different φ values (cf. SI, Fig S5 b, for HQ MAPI), and also the widths of the RC tails, inform on the single crystal's quality (41).

As a result, the XRD analysis done here allows the tracing of 1D line defects (like edge or screw dislocations), 2D planar defects (like stacking faults or grain boundaries), and/or 3D bulk defects (like voids and impurities), as these structural defects directly affect the orientation of a lattice plane, and, thus, result in changes that can be detected experimentally by comparing RCs. For 0D defects, which are reflected in RC tails, the higher dimensional defects can be viewed as including some and/or resulting from arrays of 0D defects with potential optoelectronic activity, while the higher dimensional defects can also have such activity by themselves (9, 42–44). Our measurements do not inform on the intrinsic defect levels that are thermodynamically dictated, and that, as in any ordered solid, cannot be avoided in HaPs (45, 46).

In parallel, easily applied, quick experimental methods for the preliminary assessment of the crystal quality are described. We then describe how FWHM values of the steady-state photoluminescence (PL) peak, the transient PL decay curves, and a semi quantitative analysis of the sub-bandgap surface photovoltage (SPV) response, are used to assess the optoelectronic quality of the same crystals. All the methods applied here provide bulk information; the X-ray penetration depth in HaPs is 4 μm (33), and the depth of carrier/exciton generation by supra-bandgap illumination in those materials is between 100s to a few 1000s of nanometers (see section II C below) (47, 48). In the context of detecting possible sub-bandgap defects, this volume extends deeper, because of the (much) weaker absorption of sub-bandgap photons.

This combined approach yields experimental evidence for DT in the 3D, 2D-3D, and 2D HaPs crystals, i.e., the differences in density of structural defects does NOT translate into the corresponding differences in optoelectronically active defects.

**Results and Discussion**

2D HaP $BA_2PbI_4$ (C4N1), and 2D-3D HaPs $BA_2MAPb_2I_7$ and $BA_2MA_2Pb_3I_{10}$ (C4N2 and C4N3, respectively) single crystals were prepared by using the slow cooling method (see Section S1, supplementary information, SI) (33). 3D HaP $MAPbI_3$ (MAPI) single crystals were prepared by using a modified inverse temperature crystallization method (Section S1, SI) (49, 50). The phase purity of the C4N1, C4N2, C4N3, and MAPI single crystals was verified by powder XRD on pulverized single crystals (see experimental details in Section S2, SI, and the corresponding XRD diffractograms in figure S2, SI).

**A. Structural Quality: Quantitative Evaluation**

To check the structural quality of the intact single crystals, we performed θ-2θ XRD scans directly on the crystals. The phase purity of the crystals was verified by grinding them and excluding traces of any secondary phases (see Fig. S2 in the SI). For the C4N1 crystal, diffraction peaks of the (00n) family of planes were observed (Figure 1a) (51). The narrow full width at half-maximum (FWHM) of the diffraction peaks in the θ-2θ XRD scan indicates high-quality periodicity, as expected for single crystals. To assess the degree of single crystallinity and quantify it, an omega (ω) scan is required (see Figure S1, SI); the resulting θ vs. intensity plot is called 'rocking curve' (RC). An RC analysis evaluates the perfection of a single crystal. The wider is the curve, the less perfect is the crystal and the higher is its so-called mosaicity. RCs reflect the extent of out-of-plane misorientation with regards to the specific set of planes that dominate the θ-2θ diffractograms. The extent to which those crystallographic planes deviate from being perfectly oriented with respect to one another will be manifested in the RC. The appearance of several peaks reflects the existence of sufficiently misoriented single crystalline domains so that their diffraction becomes distinguishable beyond widening the RC.

Note that, being a diffraction method, only periodic structural defects are sampled, which, in the case of HaPs are likely to be of more than one kind, due to the low free activation energies for formation and decomposition of the Pb-iodide perovskites. For C4N1 the RCs will show orientation deviations for the *{004}* family of planes. RCs for the two different C4N1 crystals (from Figure 1a) are shown in Figure 1b. The average FWHM values of the RCs of these two crystals differed widely: 0.08° vs. 0.57°, for crystals 1 and 2, respectively (the average FWHM value of each crystal represents the RC-FWHM taken at ɸ = 0°, 90°, 180°, and 270° of a specific crystal, see more in the next paragraph). Therefore, the corresponding SCs will be referred to as 'Higher Quality C4N1' (C4N1 HQ, instead of 'crystal 1'), and 'Lower Quality C4N1' (C4N1-LQ, instead of 'crystal 2'), respectively.

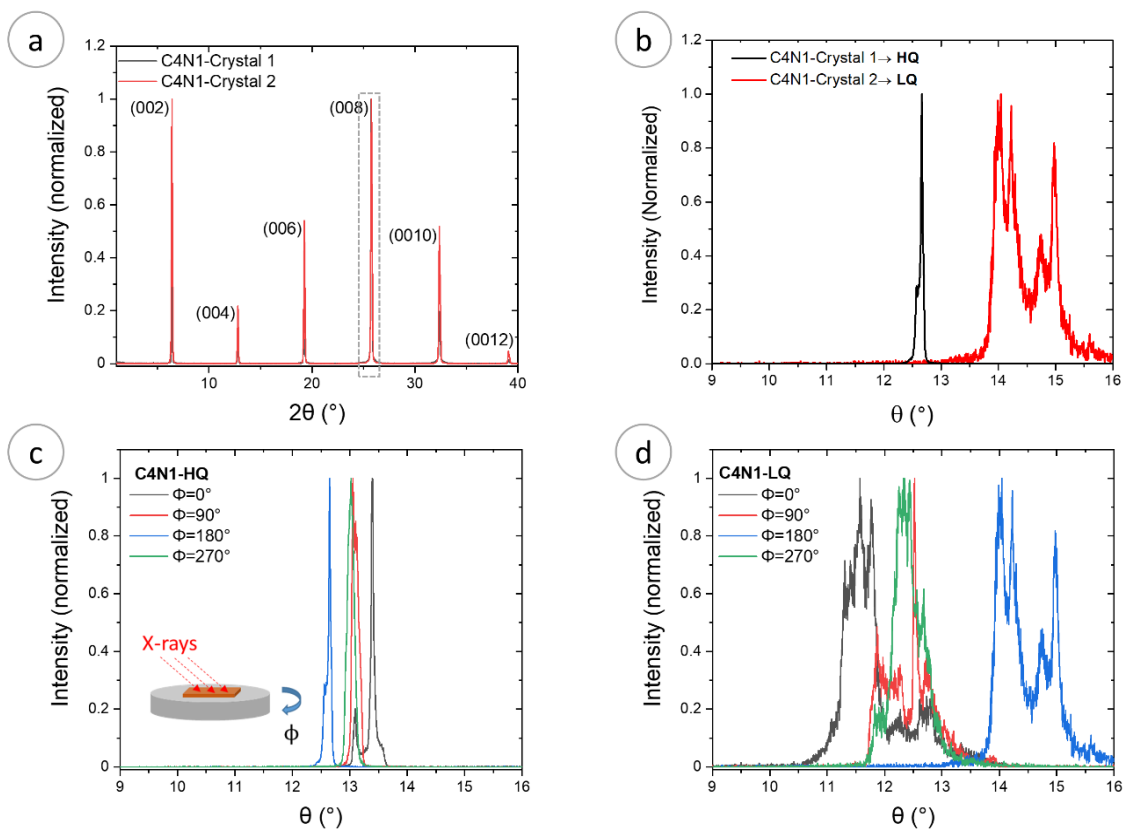

**FIG. 1.** a. θ-2θ XRD pattern for BA$_2$PbI$_4$ (C4N1), crystals 1 and 2. The peak marked with the dashed grey rectangle is the diffraction angle (25.75°), at which the rocking curves, RCs, by ω scan, are recorded. b. RC at 2θ = 25.75° and at ϕ = 180° for crystal 1 (black) with a narrow rocking curve i.e., higher quality (HQ) (average FWHM ≈ 0.08°) and for crystal 2 (red) with a wide rocking curve i.e., lower quality (LQ) (average FWHM ≈ 0.57°). c. RC for higher-quality C4N1 (C4N1-HQ) crystal performed at four different ϕ values (inset: illustration of the angle ϕ). d. RC for lower-quality C4N1 (C4N1-LQ) crystal at four different ϕ values.

From RC measurements that were performed at different ϕ values (see Figure 1c and Figure S1), we observe that the FWHM for the HQ crystal (Figure 1c) varied from 0.04° at ϕ = 0° and 180° to 0.13° at ϕ= 90° and 270°. For the C4N1-LQ crystal, dramatic differences in shape and FWHM values of the RCs were found at different ϕ values (Figure 1d). We note that due to the soft nature of the HaP single crystals, and technical limitations (laboratory, rather than synchrotron measurements), the shifts in peak positions of RCs at different ϕ values are NOT considered. The wide FWHM values and dissimilar diffraction patterns seen in the RCs of C4N1-LQ indicate significant lattice imperfections, not seen in the θ-2θ measurements. According to Dolabella et al. and Masiello et al., wider RCs indicate more out-of-plane misorientations, while variations in the RC shape and width while changing ϕ (in our case, from 0° to 90 ° to 180° to 270°) indicate in-plane misorientations (41, 52). In other words, the reason for the dominant variations between the different RCs of C4N1-LQ is that, while the stacked C4N1 crystalline flakes have similar crystallographic orientation relative to the Z axis (the direction normal to the crystal's surface, as seen from the θ-2θ XRD scan, Figure 1a), they do not have the same over-all orientation. Such can be the case when on average all the planes are orthogonal to the Z axis, but there are slight tilts between individual planes, parallel and perpendicular to the crystal's surface (the X-Y, Y-Z planes). Importantly, Kim et al. reported that for GaN epitaxial films, complex peaks (spreads) seen in RC originate from the presence of multi-domain structure (53). This indicates that (i) the HQ crystal has only slight periodic disorder compared with the LQ crystal, and that (ii) measuring the RC at different $\phi$ values is crucial for evaluating the degree of single crystallinity. Thus, even though the C4N1-HQ crystal is not a perfect single crystal, it is of much higher crystalline quality than the LQ one.

On similar lines, in the $\theta$-$2\theta$ XRD scans of 2D-3D C4N2 (Figure S3a, SI) and C4N3 crystals (Figure S4a, SI), as well as 3D MAPI crystals (Figure S5a, SI), diffraction peaks belonging to the *{h00}* (for C4N2) (54), *{0k0}* (for C4N3) (54) and *{h00}* (for MAPI) (55) families of crystallographic planes are observed with narrow FWHM (for both HQ and LQ crystals), indicating high periodicity. But, in an RC measurement ($\omega$-scan) at different $\phi$ values (like for C4N1) there is a clear difference between HQ and LQ crystals (see Figures S3b-c, S4b-c, S5b-c, SI). Overall, these results show clearly that measuring only a single crystal's $\theta$-$2\theta$ XRD scan is insufficient to gauge the structural quality of HaP single crystals. It is worth noting that the C4N1, C4N2, and C4N3 crystals, either HQ and LQ, were formed in the same crystallization system, whereas the MAPI HQ crystals were synthesized by optimizing temperature control in the inverse temperature crystallization method.

**B. Structural Quality: Qualitative Evaluation**

While the RC analysis was found to provide a quantitative evaluation of the single crystal quality, we find that a quick physical examination can qualitatively indicate about the single crystal quality. For 2D and 2D-3D single crystals, crystal quality can be assessed from simple mechanical exfoliation. After a scotch-tape exfoliation, for HQ crystals, a clean and crack-free surface is observed (C4N1-HQ, Figure 2a), while for the LQ crystals, a stepped or cracked surface is observed (C4N1-LQ, Figure 2b). Apart from that, we note that the ability to exfoliate the 2D and 2D-3D crystals is a valuable tool for exposure of a pristine surface, crucial for surface-sensitive measurements.

**Main text**

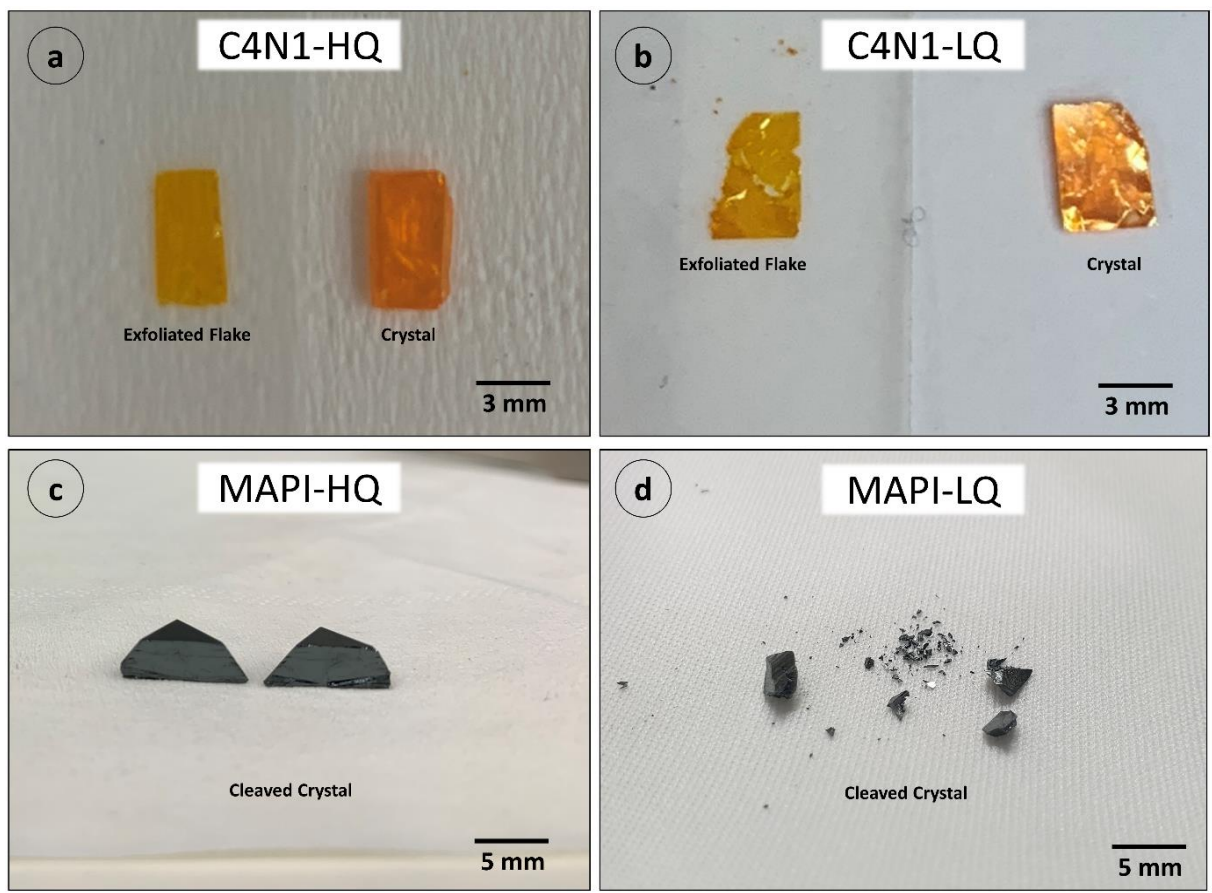

**FIG. 2**. a. Scotch-tape exfoliated C4N1-HQ (right) and its exfoliated flake (left). b. Scotch-tape exfoliated C4N1-LQ (right) and its exfoliated flake (left). c. Cleaved MAPI-HQ crystal. d. Cleaved MAPI-LQ crystal.

In addition to mechanical exfoliation, imaging the crystal with a transmission optical microscope under polarized light allows a clear and quick qualitative evaluation of crystal quality. The imaging reveals that while the transmission of polarized light for C4N1-HQ is uniform throughout the sample (Figure S6b, SI), it appears nonuniform for C4N1-LQ (Figure S6d, SI). This is because due to anisotropy, the polarized light travels at different velocities through the material and takes different pathways depending on the light beam's orientation, relative to the crystallographic orientation of the crystal. In this method, which is used widely in mineralogy and gemology (56), the resulting color contrast for polarized light transmission reflects differently oriented volumes in the crystal. Here the contrasting features across the C4N1-LQ crystal confirm that it is more heterogeneous than the C4N1-HQ crystal.

For 3D MAPI crystals, a quick qualitative indication can be observed by cleaving the crystal parallel to a crystal plane. If a clean cleavage along a particular crystallographic plane is achieved (which leads to the exposure of mirror-like crystal faces of the two freshly exposed surfaces), it indicates a high-quality single crystal, corresponding to the HQ single crystal (Figure 2c and Supplementary Video 1). Instead, if the crystal breaks into many fragments while cleaving, it indicates that this is a LQ single crystal (Figure 2d and Supplementary Video 2). While this physical examination of the 3D crystals is destructive (to some degree, depending on the quality of the parts after cleavage), for 2D crystals the scotch tape exfoliation examination can be considered (almost) non-destructive. Thus, for 2D crystals, this physical examination can be easily applied.

**C. Optoelectronic Quality: Qualitative and Semi-quantitative Evaluation**

Now that we have established the structural differences between HQ and LQ crystals of various compositions, we further look at two properties that can show evidence for opto-electronically active defects in the C4N1 (2D), C4N2 (2D-3D) and MAPI (3D) crystals: photoluminescence (PL), and surface photovoltage (SPV) measurements. Our hypothesis is that, in the absence of DT, for a given composition and structure, a crystal of higher structural quality (i.e., HQ) will have: 1) a narrower FWHM of the main PL emission peak, indicating a smaller density of band tail states, or, more generally, of shallow defect states near the band edges, 2) longer carrier lifetimes, and 3) smaller sub-bandgap SPV response (indicating a lower density of deep defect states within the forbidden gap) than those of a LQ crystal, i.e., one of poorer structural quality. The reasoning is that, as for common semiconductors, structural disorder implies structural defects, which can be traced to 0D point defects (where a single point in the actual lattice differs from the perfect lattice structure) (57). The electronic activity of these defects leads, in turn, to charge/discharge states with energy levels between the band edges, i.e., within the forbidden gap (bandgap). Mostly theoretical, and some experimental (for 2D defects) data have shown that structural defects in HaPs are effectively the most attractive channels for the migration of point defects and that the 0D defects preferably form at 2D defects, via grain boundaries and surfaces (58, 59). The presence of such sub-bandgap states will affect the (opto)electronic quality of the crystal. *Direct* experimental identification of 0D structural point defects (i.e., different from inferring their presence from interpreting spectra or diffraction data), is non-trivial and has, to the best of our knowledge not been attempted for HaPs at levels below full atomic percentages of foreign material (e.g., Bi "doping") (58, 59). Specifically, XRD-based evidence for a ~0.2% (~1 pm) $CsPbI_2Br$ lattice shrinking upon doping with Zn(II) or Sb(III) at 1 at.% Pb was shown on polycrystalline films with at least a few 100s of nm grain size (to exclude surface doping) (60). Another study of Sm(II) (0.7-3.5 at.% Pb) doping of MAPI showed that bulk dopant incorporation led to 0.5 pm increase in lattice parameter for all Sm levels, suggesting Sm(II)→ Sm(III) oxidation (61). The doping of $FAPbI_3$ with Nd(III) (5 at.% Pb), ended up in ~1.4% change in lattice parameter, with similar results for doping with Ca(II) and Na(I) (62). Furthermore, in one case, doping of $MAPbBr_3$ single crystals by 1 at.% Pb of Bi(III) gave a 1000 fold increase in (n-type) conductivity, with ~0.01 % lattice parameter decrease, but some increase in diffraction peak widths (63). The presence of such sub-bandgap states will affect the (opto)electronic quality of the crystal.

We find that for a given composition, the FWHM of the PL emission peak is essentially not affected by the structural quality of the crystal (Figure S7, SI). This finding already suggests that with respect to tail states, no correlation is found between the presence of structural defects and the emergence of PL-detectable densities of shallow defects near the band edges.

Although time-resolved PL, TRPL, is widely used for assessing the optoelectronic quality of HaP samples, accumulating experimental evidence suggests that, while extracting a single PL lifetime value may be useful for comparing different samples in the same lab (as we tried to do in this study) (64), absolute PL lifetime values should be viewed with care. The reason is that they can vary significantly between laboratories, due to factors such as sample preparation, surface passivation, excitation parameters, measurement atmosphere, and the models used for the data analyses (65–70).

This point is further supported by the results of a recent meta-analysis by Yuan et al., who studied TRPL data measured on HaPs from many published papers (70). They find that the extraction of lifetimes is very problematic when it comes to HaPs, and that accordingly, the same holds for conclusions based on those values. In addition, surface states and crystals' geometry were shown to have a primary influence on TRPL results of HaPs (71–75).

Still, TRPL data of HQ and LQ single crystals of 2D, 2D-3D, and 3D HaPs were collected (see figures S8 and S9, as well as table S1 in the SI) to evaluate if the results can be used as comparative tool. We used two different systems for this, a home-built and a commercial one with comparable results. To remove some of the problems inherent in TRPL lifetime measurements the results that we report here were measured by the commercial system, i.e., the same instrument, for all samples, and using identical excitation conditions for any given set of HQ and LQ HaP samples. We used a low excitation intensity, with photon pulse flux of ~2.5 x $10^9$ photons/cm$^2$ (5 x $10^{10}$ photons/cm$^2$ for the home-built system). As a result of the overlap between the times of the C4N1 TRPL decay and that of the excitation beam (system response), those data cannot be analyzed meaningfully (see figure S8a). The C4N2 samples were not measured because of technical problems.

We tried to obtain values for lifetimes by fitting the decay data for each of the C4N3 and MAPI single crystals, using both mono- and bi-exponential functions (see SI, Figs. S8 b, d and c, e). Neither mono- nor bi-exponential decay functions produced a good fit for the collected data. With this *caveat* in mind, the values for the HQ samples are about half those for the LQ ones (see Table 1 and SI Table S1). Slow decay time, $\tau_2$, values, obtained from using the bi-exponential function, range for both HQ and LQ crystals from tens to over a hundred ns, comparable to those reported in the literature for single crystals of MAPI and C4N3 (see Fig. S9, SI) (64, 72–74, 76–79).

Table 1. Results of PL decay fitting attempts to biexponential functions

| Composition | Crystal-Quality | PL lifetimes (nsec) $\tau_1 / \tau_2$ * |
|---|---|---|
| **C4N3 (2D-3D)** | HQ | 1.4 / 14 |
| | LQ | 2.5 / 24 |
| **MAPI (3D)** | HQ | 7.5 / 46 |
| | LQ | 15 / 115 |

\* See *caveat* re. the bi-exponential fit to the TRPL decay curves in the text. The raw TRPL data are shown in SI Figs. S8 and S9 and given in more detail in Table S1

Furthermore, PL quantum yield measurements, which theoretically would be a good measure for the evaluation of the SCs' qualities, are commonly not used for HaP single crystals as indicator for crystal quality (72–74, 76, 77). Probably, this is, at least in part because the systems used for this measurement rely on the assumption that the sample is optically thin (negligibly reabsorbing the emitted light), while such is not the case for single crystals and for most PV-quality thin films of HaPs, materials that are known for their high absorption coefficients.

For SPV, we focus solely on sub-bandgap SPV photons and the sample's response. SPV is a technique that, in contrast to the more commonly used PL, does not rely on radiative recombination processes. An SPV signal is observed in cases where the photogeneration of charge carriers results in charge carrier separation in space. Because HaPs are high-quality (i.e., low optoelectronically active defect density) semiconductors, detecting defect-related signals poses a challenge. To meet that challenge, light-modulated SPV measurements in combination with lock-in detection were employed, in order to significantly increase the sensitivity and S/N ratio. In addition to the challenge of detection, there is the as yet unsolved issue of quantifying the defect densities that give rise to the SPV signal. In our case, we propose a **<u>comparative</u>** semi-quantitative assessment of the defect densities by integrating the area below the SPV curve, at energies less than the bandgap (see more about this in the SI, section S4). Such an approximation was previously employed to compare defect densities in Halide Perovskites from photoemission measurements by Lee et al (80).

Optical absorption coefficients for photon energies below the bandgap energy are orders of magnitude lower than those above (at higher photon energies than the) bandgap, at 450 nm for the 2D and 2D-3D samples, or at 640 nm for MAPI, the excitation wavelengths that we used for the TRPL experiments with home-built system (see SI, section S2). This implies that sub-bandgap absorption, and, therefore, any possible carrier generation/release occurs orders of magnitude deeper into the crystals than carrier generation due to photons with energies above the bandgap (which will be limited to a few 100 nm). For sub-bandgap absorption in MAPI, any carrier generation/ release (from traps) will occur up to $10^4$ nm, and for C4N1 up to $3 \times 10^3$ nm into the sample (48). Therefore, with sub bandgap photon excitation the change in contact potential difference, ΔCPD, due to photon absorption, i.e., the SPV, reflects also what happens in the bulk of the crystals (47, 81, 82).

As extracting defect densities from SPV spectra is still a challenging task (82), we approximate the relative density of defects in the sample volume that is measured (which is constant for all samples) by integrating the area under the curve of the sub bandgap SPV response (above the background signal; cf. figure 3 and see SI, section 4 for further calculation details). We note that, sub bandgap SPV responses with and without crystal surface exfoliation prior to the measurement are almost similar (see Figure S10, SI, for C4N1-HQ crystal).

From comparing the integration results (between HQ and LQ crystals), we find that, in line with the PL results, there are NO consistent differences between their sub-bandgap SPV responses. For C4N1, C4N2, C4N3 and MAPI, the integrated SPV spectral areas for the HQ vs. LQ are either the same or almost identical, as the minor differences between them are close to 3 times the experimental error (Figure 3, see also figure S12 in which we show that this conclusion is based on spectra of multiple samples). Furthermore, these differences in SPV between HQ and LQ crystals are indeed minor, because, using the same technique, a clear and significant change in SPV was reported for a systematic and slight change in extrinsic defect density by doping HaP with Bi. (83) Therefore, the results presented here that indicate the DT of intrinsic structural defects of Pb-HaPs are genuinely factual ones. Importantly, the type of sub-bandgap SPV in the fixed capacitor arrangement used in this study is one of the most sensitive contact-less techniques available today for detecting signals related to opto-electronically active defects (84).

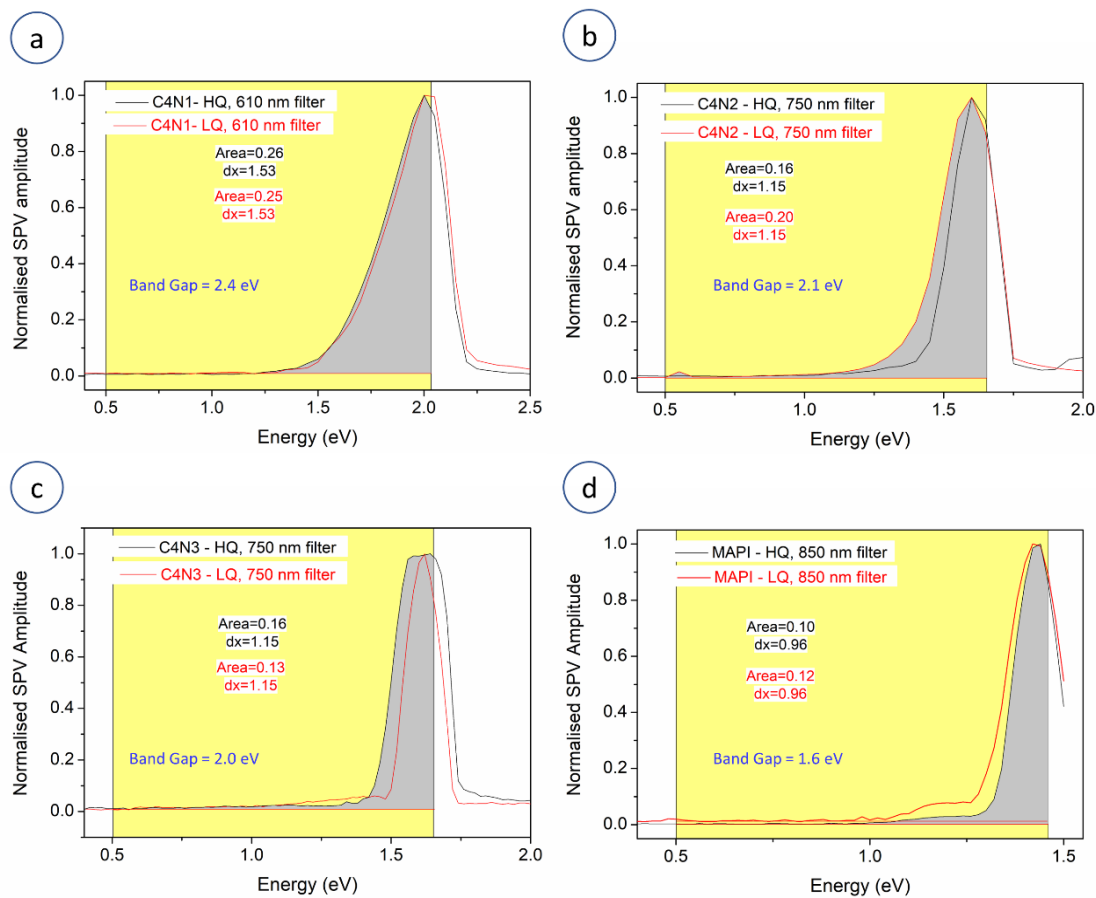

**FIG. 3.** Modulated sub-bandgap SPV spectra for HQ and LQ crystals with appropriate long pass filters a. C4N1 b. C4N2 c. C4N3 d. MAPI. Area represents the area under the curve (grey zone), calculated between the limits of the rectangle edges, (dx) (yellow zone), with the baseline (lower limit on y scale) set at the SPV noise level. For clarity only one such grey zone is shown for each pair of (black and red) spectra. On the x (= photon energy) axis the upper limit corresponds to the cut-on energy of the respective long-pass filter used for each composition and the lower limit is around 0.5 eV, the photon energy, where any SPV signal disappears in the noise level. In the legends dx is the difference between the low and high photon energy limits, used to calculate the Area. See SI, section 4 for further error calculation details.

### D. Comparison of Structural and Optoelectronic Quality

The results show that there is no correlation between structural defects and opto-electronically active defects in the 2D, 2D-3D, 3D HaPs that we studied (cf. Table 2 SPV integrated areas and FWHM of RCs). It is this absence that provides straightforward experimental evidence for defect tolerance, DT. This indicates that DT is related to the inorganic lattice, i.e., the sheet(s) of $PbI_6$ octahedra, a common component for all the HaPs studied in this work.

Our results can be explained by the following reasons that are connected to each other:
1. The relative softness of the perovskite lattice has been used to invoke formation of large polarons that would then result in efficient columbic screening of structural defects, to explain the optoelectronic quality of HaPs (85, 86). If correct, it could result in the lack of measurable optoelectronic effects that reflect electron-hole (non-radiative) recombination due to the relative increase in structural defects (from HQ to LQ crystals) (27, 28, 87, 88). However, some reproducible experimental results are clearly inconsistent with this view (89).2. More fundamentally, the low deformation energy of Pb-HaPs, as expressed also in their soft lattice, could allow the material to tolerate structural defect-induced strain fields, which will mitigate the effects of the structural defects on the Pb-HaP optoelectronic characteristics (45, 90).
3. In addition, the dynamic local disorder resulting in a dynamic electronic landscape in HaPs, possibly masks the effects of the densities of structural defects at doping density levels (24, 91, 92).

Following our experimental observations, the plausible mechanism for DT in Pb-HaPs seems to be associated with reasons -2- and -3-, the low lattice deformation energy and dynamic local disorder of Pb-HaPs, which is in line with the conclusions drawn from the computational studies that were reported earlier (24, 28). We note that, while our experimental results do not yet include all-inorganic 3D HaPs, the apparently central role of the Pb-halide lattice, suggests that all Pb HaPs, will show some degree of DT (28). On a side note, the exceptional self-healing of Pb-HaPs is also ascribed to one or more of the above-mentioned features, in addition to hydrogen bonding between halides and organic cation for hybrid Pb-HaPs.

**TABLE 2.** Comparison of structural and optoelectronic quality of the four HaP crystal types. The FWHM values of the RC were extracted from the data presented in figures S11b-c, S3b-c, S4b-c, and S5b-c for C4N1, C4N2, C4N3 and MAPI, respectively. The SPV values were extracted from the data presented in figure 3 while considering the bandgap energies of C4N1, C4N2, C4N3 and MAPI. The error of the integrated SPV values is in the third digit after the decimal point; thus, it is negligible here; for more details see Section S4 in the SI.

| Composition | Crystal-Quality | RC-FWHM (a.u.) | SPV (Area) (a.u.) |
|---|---|---|---|
| C4N1 (2D) | HQ | 0.18 | 0.26 |
|  | LQ | 1.02 | 0.25 |
| C4N2 (2D-3D) | HQ | 0.08 | 0.16 |
|  | LQ | 0.78 | 0.20 |
| C4N3 (2D-3D) | HQ | 0.10 | 0.16 |
|  | LQ | 0.43 | 0.13 |
| MAPI (3D) | HQ | 0.04 | 0.10 |
|  | LQ | 5.68 | 0.12 |

As noted elsewhere (37), the effect of DT on materials (and, thus, on device) stability, must reach a limit if defects accumulate. SH capability keeps the defect density at a "tolerable" level. By having both DT and SH, DT will be an important stabilizing effect in those HaPs that have a slower rate of SH (e.g., CsPbBr$_3$ compared to FAPbBr$_3$) by enabling a wide range of SH rates to be effective for material stability.

**Conclusion**

Overall, our results of rocking curve XRD (RC-XRD) analysis, as a measure of structural quality, and PL and sub-band gap SPV response as a measure of optoelectronic quality, show a lack of correlation between the RC-XRD and the PL/SPV results, for the 2D (C4N1), 2D-3D crystals (C4N2, C4N3) and 3D MAPI crystals. We suggest that the reason for this is the soft and highly dynamic (around a well-defined average) structure of the HaP lattice, especially the inorganic PbI$_6$ octahedra component, the structural backbone of the lead HaPs studied in this work. Such dynamic lattice affects the electronic landscape, to overshadow or compensate electronic effects to an extent that any effects are below the detection limits of the highly sensitive technique/s employed in this work. This result also implies that for Pb-HaPs, the structural quality of crystals is, as far as we can measure by the means employed here, a poor predictor for optoelectronically active defects. We purposely focus on macroscopic-sized single crystals, instead of on polycrystalline films with sub-mm grain sizes (i.e., high surface/volume ratios). Thus, our results do not inform on the relation between surface defects and their effect on the optoelectronic properties of HaPs of various dimensionalities, which requires also the use of specifically surface- vs. bulk-sensitive characterization tools.

We posit that the absence of correlation between evidence for defects from structural experiments on the one hand and optoelectronic ones on the other hand provides the first clear experimental evidence for defect tolerance in Pb-iodide HaPs, as DT appears the most obvious (Ockham's razor) interpretative explanation. As such, we look forward to results from high-level computations, including the lattice dynamics, and from higher level (synchrotron-based) RC measurements, possibly complemented by other measurements that can match the sensitivity of SPV, to test our interpretation, and with experimentally-testable predictive power. The present results add to our level of understanding of the fascinating science of Pb-HaPs. The results can furthermore be viewed also in light of our earlier self-healing studies on 2D, 2D-3D, and 3D HaPs. A clear direction for future fundamental experimental work is to design experiments that can test the possible co-existence of defect tolerance and self-healing in HaPs.

In summary, defect tolerance in Pb-HaPs suggests that the dynamic columbic screening in intrinsically "soft" and polarizable materials provides a design principle to achieve some co-existence of structural imperfections and close-to-optoelectronic perfection in materials. As such, DT can bridge any time lag between damage and self-healing to minimize effects of imperfections, to make materials more sustainable.

**ACKNOWLEDGEMENTS**

NPJ and SA thank Dr. Davide Raffele Cerrati for providing the MAPI-LQ crystals, Dr. Sujit Kumar for help with PL measurements, and Dr. Anna Kossoy for assistance in the XRD measurements. IL, SA and DC thank Prof. Lioz Etgar and Tal Binyamin for their help with the TRPL measurements. NPJ acknowledges funding from the European Union's Horizon 2020 MSCA Innovative Training Network MAESTRO under grant agreement no. 764787. At Bar-Ilan University work was supported also by the Israel Ministry of Energy as part of the Solar ERAnet PerDry consortium. At the Weizmann Institute of Science (DC) the work received support from the Minerva Centre for Self-Repairing Systems for Energy & Sustainability and the Sustainability and Energy Research Initiative, SAERI. IL thanks the AiF project (ZIM-KK5085302DF0) for financial support and Dr. Thomas Dittrich for fruitful conversations.

**Declaration of Interest:**
The authors declare no competing interests.

**Author contributions:**
NPJ, SA, and DC designed the research; NPJ, IL, YF and SA performed the experiments; NPJ, IL, YF, SA, and DC analyzed the data; NPJ, IL, SA, and DC wrote the manuscript. GH contributed to the data analysis and manuscript review. DC supervised this research work. All authors discussed and finalised the manuscript.

**References**

1. W. Zhang, G. E. Eperon, H. J. Snaith, Metal halide perovskites for energy applications. *Nat. Energy 2016 16* **1**, 1–8 (2016).
2. J. S. Manser, J. A. Christians, P. V. Kamat, Intriguing Optoelectronic Properties of Metal Halide Perovskites. *Chem. Rev.* **116**, 12956–13008 (2016).
3. H. Kim, J. S. Han, J. Choi, S. Y. Kim, H. W. Jang, Halide Perovskites for Applications beyond Photovoltaics. *Small Methods* **2**, 1700310 (2018).
4. L. Chouhan, S. Ghimire, C. Subrahmanyam, T. Miyasaka, V. Biju, Synthesis, optoelectronic properties and applications of halide perovskites. *Chem. Soc. Rev.* **49**, 2869–2885 (2020).
5. T. Miyasaka, Lead Halide Perovskites in Thin Film Photovoltaics: Background and Perspectives. *https://doi.org/10.1246/bcsj.20180071* **91**, 1058–1068 (2018).
6. L. A. Muscarella, B. Ehrler, The influence of strain on phase stability in mixed-halide perovskites. *Joule* **6**, 2016–2031 (2022).
7. Y. Hu, *et al.*, Highly efficient flexible solar cells based on a room-temperature processed inorganic perovskite. *J. Mater. Chem. A* **6**, 20365–20373 (2018).
8. Y. M. Wang, *et al.*, High-Efficiency Flexible Solar Cells Based on Organometal Halide Perovskites. *Adv. Mater.* **28**, 4532–4540 (2016).

9. H. J. Queisser, E. E. Haller, Defects in semiconductors: some fatal, some vital. *Science* **281**, 945–950 (1998).
10. A. Zakutayev, *et al.*, Defect tolerant semiconductors for solar energy conversion. *J. Phys. Chem. Lett.* **5**, 1117–1125 (2014).
11. S. Zhang, S. H. Wei, A. Zunger, H. Katayama-Yoshida, Defect physics of the $CuInSe_2$ chalcopyrite semiconductor. *Phys. Rev. B* **57**, 9642–9656 (1998).
12. A. K. Ramdas, S. Rodriguez, Spectroscopy of the solid-state analogues of the hydrogen atom: donors and acceptors in semiconductors. *Reports Prog. Phys.* **44**, 1297 (1981).
13. W.-J. Yin, T. Shi, Y. Yan, Unique Properties of Halide Perovskites as Possible Origins of the Superior Solar Cell Performance. *Adv. Mater.* **26**, 4653–4658 (2014).
14. C. He, X. Liu, The rise of halide perovskite semiconductors. *Light. Sci. Appl.* **12**, 2047–7538 (2023).
15. T. M. Brenner, D. A. Egger, L. Kronik, G. Hodes, D. Cahen, Hybrid organic—inorganic perovskites: low-cost semiconductors with intriguing charge-transport properties. *Nat. Rev. Mater. 2016 11* **1**, 1–16 (2016).
16. Y. -T. Huang, *et al.*, Perovskite-inspired materials for photovoltaics and beyond—from design to devices. *Nanotechnology* **32**, 132004 (2021).
17. X. Zhang, M. E. Turiansky, C. G. Van de Walle, Correctly Assessing Defect Tolerance in Halide Perovskites. *J. Phys. Chem. C* **124**, 6022–6027 (2020).
18. J. S. Park, S. Kim, Z. Xie, A. Walsh, Point defect engineering in thin-film solar cells. *Nat. Rev. Mater. 2018 37* **3**, 194–210 (2018).
19. W. J. Yin, T. Shi, Y. Yan, Unusual defect physics in $CH_3NH_3PbI_3$ perovskite solar cell absorber. *Appl. Phys. Lett.* **104**, 63903 (2014).
20. J. Kim, S.-H. Lee, J. H. Lee, K.-H. Hong, The Role of Intrinsic Defects in Methylammonium Lead Iodide Perovskite. *J. Phys. Chem. Lett.* **5**, 1312–1317 (2014).
21. J. Kim, C. H. Chung, K. H. Hong, Understanding of the formation of shallow level defects from the intrinsic defects of lead tri-halide perovskites. *Phys. Chem. Chem. Phys.* **18**, 27143–27147 (2016).
22. R. C. Kurchin, P. Gorai, T. Buonassisi, V. Stevanović, Structural and Chemical Features Giving Rise to Defect Tolerance of Binary Semiconductors. *Chem. Mater.* **30**, 5583–5592 (2018).
23. X. Zhang, M. E. Turiansky, J. X. Shen, C. G. Van De Walle, Defect tolerance in halide perovskites: A first-principles perspective. *J. Appl. Phys.* **131** (2022).
24. A. V. Cohen, D. A. Egger, A. M. Rappe, L. Kronik, Breakdown of the Static Picture of Defect Energetics in Halide Perovskites: The Case of the Br Vacancy in $CsPbBr_3$. *J. Phys. Chem. Lett.* **10**, 4490–4498 (2019).
25. B. Wang, *et al.*, Electron-Volt Fluctuation of Defect Levels in Metal Halide Perovskites on a 100 ps Time Scale. *J. Phys. Chem. Lett.* **13**, 5946–5952 (2022).
26. T. Kirchartz, T. Markvart, U. Rau, D. A. Egger, Impact of Small Phonon Energies on the Charge-Carrier Lifetimes in Metal-Halide Perovskites. *J. Phys. Chem. Lett.* **9**, 939–946 (2018).
27. W. Chu, Q. Zheng, O. V. Prezhdo, J. Zhao, W. A. Saidi, Low-frequency lattice phonons in halide perovskites explain high defect tolerance toward electron-hole recombination. *Sci. Adv.* **6**, 7453–7467 (2020).
28. W. Chu, W. A. Saidi, J. Zhao, O. V. Prezhdo, Soft Lattice and Defect Covalency Rationalize Tolerance of β-$CsPbI_3$ Perovskite Solar Cells to Native Defects. *Angew. Chemie Int. Ed.* **59**, 6435–6441 (2020).

29. W. Liu, *et al.*, Mapping Trap Dynamics in a CsPbBr$_3$ Single-Crystal Microplate by Ultrafast Photoemission Electron Microscopy. *Nano Lett.* **21**, 2932–2938 (2021).
30. D. R. Ceratti, *et al.*, Self-Healing Inside APbBr$_3$ Halide Perovskite Crystals. *Adv. Mater.* **30**, 1706273 (2018).
31. W. Nie, *et al.*, Light-activated photocurrent degradation and self-healing in perovskite solar cells. *Nat. Commun. 2016 71* **7**, 1–9 (2016).
32. D. R. Ceratti, *et al.*, The pursuit of stability in halide perovskites: the monovalent cation and the key for surface and bulk self-healing. *Mater. Horizons* **8**, 1570–1586 (2021).
33. S. Aharon, *et al.*, 2D Pb-Halide Perovskites Can Self-Heal Photodamage Better than 3D Ones. *Adv. Funct. Mater.* **32**, 2113354 (2022).
34. B. P. Finkenauer, Akriti, K. Ma, L. Dou, Degradation and Self-Healing in Perovskite Solar Cells. *ACS Appl. Mater. Interfaces* **14**, 24073–24088 (2022).
35. M. Holland, *et al.*, Metal Halide Perovskites Demonstrate Radiation Hardness and Defect Healing in Vacuum. *ACS Appl. Mater. Interfaces* **14**, 9352–9362 (2022).
36. P. Singh, *et al.*, A-Site Cation Dependence of Self-Healing in Polycrystalline APbI$_3$ Perovskite Films. *ACS Energy Lett.* **8**, 2447–2455 (2023).
37. D. Cahen, L. Kronik, G. Hodes, Are Defects in Lead-Halide Perovskites Healed, Tolerated, or Both? *ACS Energy Lett.* **6**, 4108–4114 (2021).
38. A. Walsh, A. Zunger, Instilling defect tolerance in new compounds. *Nat. Mater. 2017 1610* **16**, 964–967 (2017).
39. R. E. Brandt, V. Stevanović, D. S. Ginley, T. Buonassisi, Identifying defect-tolerant semiconductors with high minority-carrier lifetimes: Beyond hybrid lead halide perovskites. *MRS Commun.* **5**, 265–275 (2015).
40. G. Xing, *et al.*, Low-temperature solution-processed wavelength-tunable perovskites for lasing. *Nat. Mater. 2014 135* **13**, 476–480 (2014).
41. S. Dolabella, *et al.*, Lattice Strain and Defects Analysis in Nanostructured Semiconductor Materials and Devices by High-Resolution X-Ray Diffraction: Theoretical and Practical Aspects. *Small Methods* **6**, 2100932 (2022).
42. E. B. Yakimov, Dislocation-Point Defect Interaction Effect on Local Electrical Properties of Semiconductors. *J. Phys. III* **7**, 2293–2307 (1997).
43. H. S. Leipner, J. Schreiber, H. Uniewski, S. Hildebrandt, Dislocation luminescence in CdTe. *Scanning Microsc.* **12**, 149–160 (1998).
44. P. G. Callahan, B. B. Haidet, D. Jung, G. G. E. Seward, K. Mukherjee, Direct observation of recombination-enhanced dislocation glide in heteroepitaxial GaAs on silicon. *Phys. Rev. Mater.* **2**, 081601 (2018).
45. Y. Rakita, D. Cahen, G. Hodes, Between Structure and Performance in Halide Perovskites for Photovoltaic Applications: the Role of Defects [arXiv:1809.10949v1](arXiv:1809.10949v1) *[cond-mat.mtrl-sci]* (2018).
46. Y. Rakita, I. Lubomirsky, D. Cahen, When defects become 'dynamic': halide perovskites: a new window on materials? *Mater. Horizons* **6**, 1297–1305 (2019).
47. S. D. Stranks, *et al.*, Electron-hole diffusion lengths exceeding 1 micrometer in an organometal trihalide perovskite absorber. *Science (80-. ).* **342**, 341–344 (2013).
48. K. Matsuishi, T. Ishihara, S. Onari, Y. H. Chang, C. H. Park, Optical properties and structural phase transitions of lead-halide based inorganic–organic 3D and 2D perovskite semiconductors under high pressure. *Phys. status solidi* **241**, 3328–3333 (2004).
49. M. I. Saidaminov, A. L. Abdelhady, G. Maculan, O. M. Bakr, Retrograde solubility of

formamidinium and methylammonium lead halide perovskites enabling rapid single crystal growth. *Chem. Commun.* **51**, 17658–17661 (2015).

50. N. P. Jasti, *et al.*, The Saga of Water and Halide Perovskites: Evidence of Water in Methylammonium Lead Tri-Iodide. *Adv. Funct. Mater.* **32**, 2204283 (2022).
51. M. Menahem, *et al.*, Strongly Anharmonic Octahedral Tilting in Two-Dimensional Hybrid Halide Perovskites. *ACS Nano* **15**, 10153–10162 (2021).
52. F. Masiello, *et al.*, Rocking curve measurements revisited. *urn:issn:1600-5767* **47**, 1304–1314 (2014).
53. J. Kim, *et al.*, Anisotropic mosaicity and lattice-plane twisting of an m-plane GaN homoepitaxial layer. *CrystEngComm* **21**, 4036–4041 (2019).
54. W. Paritmongkol, *et al.*, Synthetic Variation and Structural Trends in Layered Two-Dimensional Alkylammonium Lead Halide Perovskites. *Chem. Mater.* **31**, 5592–5607 (2019).
55. J. Ding, *et al.*, Design Growth of MAPbI$_3$ Single Crystal with (220) Facets Exposed and Its Superior Optoelectronic Properties. *J. Phys. Chem. Lett.* **9**, 216–221 (2018).
56. F. D. Bloss, *An Introduction to the Methods of Optical Crystallography* (Holt, Rinehart and Winston, 1961).
57. D. Moia, J. Maier, Ionic and electronic energy diagrams for hybrid perovskite solar cells. *Mater. Horizons* (2023) https:/doi.org/10.1039/D2MH01569B.
58. Y. Zhou, I. Poli, D. Meggiolaro, F. De Angelis, A. Petrozza, Defect activity in metal halide perovskites with wide and narrow bandgap. *Nat. Rev. Mater. 2021 611* **6**, 986–1002 (2021).
59. J. S. Park, J. Calbo, Y. K. Jung, L. D. Whalley, A. Walsh, Accumulation of Deep Traps at Grain Boundaries in Halide Perovskites. *ACS Energy Lett.* **4**, 1321–1327 (2019).
60. L. Hu, *et al.*, Valence-Regulated Metal Doping of Mixed-Halide Perovskites to Modulate Phase Segregation and Solar Cell Performance. *ACS Energy Lett.* **7**, 4150–4160 (2022).
61. Z. Molenda, *et al.*, Redox-active ions unlock substitutional doping in halide perovskites. *Mater. Horizons* **10**, 2845–2853 (2023).
62. Y. Zhao, *et al.*, Suppressing ion migration in metal halide perovskite via interstitial doping with a trace amount of multivalent cations. *Nat. Mater.* **21**, 1396–1402 (2022).
63. W. Tang, *et al.*, Substitutional doping of hybrid organic–inorganic perovskite crystals for thermoelectrics. *J. Mater. Chem. A* **8**, 13594–13599 (2020).
64. L. Kong, *et al.*, Simultaneous band-gap narrowing and carrier-lifetime prolongation of organic-inorganic trihalide perovskites. *Proc. Natl. Acad. Sci. U. S. A.* **113**, 8910–8915 (2016).
65. I. Levine, *et al.*, Can we use time-resolved measurements to get steady-state transport data for halide perovskites? *J. Appl. Phys.* **124** (2018).
66. F. Staub, *et al.*, Beyond Bulk Lifetimes: Insights into Lead Halide Perovskite Films from Time-Resolved Photoluminescence. *Phys. Rev. Appl.* **6**, 044017 (2016).
67. E. V. Peán, S. Dimitrov, C. S. De Castro, M. L. Davies, Interpreting time-resolved photoluminescence of perovskite materials. *Phys. Chem. Chem. Phys.* **22**, 28345–28358 (2020).
68. T. Yamada, *et al.*, Fast Free-Carrier Diffusion in CH$_3$NH$_3$PbBr$_3$ Single Crystals Revealed by Time-Resolved One- and Two-Photon Excitation Photoluminescence Spectroscopy. *Adv. Electron. Mater.* **2**, 1500290 (2016).
69. A. Capitaine, B. Sciacca, Monocrystalline Methylammonium Lead Halide Perovskite Materials for Photovoltaics. *Adv. Mater.* **33**, 2102588 (2021).

70. Y. Yuan, *et al.*, Shallow defects and variable photoluminescence decay times up to 280 µs in triple-cation perovskites. *Nat. Mater. 2024*, 1–7 (2024).
71. H. H. Fang, *et al.*, Ultrahigh sensitivity of methylammonium lead tribromide perovskite single crystals to environmental gases. *Sci. Adv.* **2** (2016).
72. L. Zhao, *et al.*, Surface-defect-passivation-enabled near-unity charge collection efficiency in bromide-based perovskite gamma-ray spectrum devices. *Nat. Photonics 2024*, 1–8 (2024).
73. R. Chuliá-Jordán, E. J. Juarez-Perez, Short Photoluminescence Lifetimes Linked to Crystallite Dimensions, Connectivity, and Perovskite Crystal Phases. *J. Phys. Chem. C* **126**, 3466–3474 (2022).
74. Y. Song, *et al.*, Efficient lateral-structure perovskite single crystal solar cells with high operational stability. *Nat. Commun. 2020 111* **11**, 1–8 (2020).
75. H. Zhu, *et al.*, Organic Cations Might Not Be Essential to the Remarkable Properties of Band Edge Carriers in Lead Halide Perovskites. *Adv. Mater.* **29**, 1603072 (2017).
76. P. K. Nayak, *et al.*, Mechanism for rapid growth of organic–inorganic halide perovskite crystals. *Nat. Commun. 2016 71* **7**, 1–8 (2016).
77. D. Shi, *et al.*, Low trap-state density and long carrier diffusion in organolead trihalide perovskite single crystals. *Science (80-. ).* **347** (2015).
78. Y. Yamada, *et al.*, Dynamic Optical Properties of $CH_3NH_3PbI_3$ Single Crystals As Revealed by One- and Two-Photon Excited Photoluminescence Measurements. *J. Am. Chem. Soc.* **137**, 10456–10459 (2015).
79. Y. Bi, *et al.*, Charge Carrier Lifetimes Exceeding 15 µs in Methylammonium Lead Iodide Single Crystals. *J. Phys. Chem. Lett.* **7**, 923–928 (2016).
80. S. Lee, *et al.*, Control of Interface Defects for Efficient and Stable Quasi-2D Perovskite Light-Emitting Diodes Using Nickel Oxide Hole Injection Layer. *Adv. Sci.* **5**, 1801350 (2018).
81. W. Zhou, *et al.*, Electronic and optical absorption properties of organic–inorganic perovskites as influenced by different long-chain diamine molecules: first-principles calculations. *RSC Adv.* **9**, 14718–14726 (2019).
82. L. Kronik, Y. Shapira, Surface photovoltage phenomena: theory, experiment, and applications. *Surf. Sci. Rep.* **37**, 1–206 (1999).
83. M. Yavari, *et al.*, How far does the defect tolerance of lead-halide perovskites range? The example of Bi impurities introducing efficient recombination centers. *J. Mater. Chem. A* **7**, 23838–23853 (2019).
84. I. Levine, *et al.*, Deep Defect States in Wide-Band-Gap $ABX_3$ Halide Perovskites. *ACS Energy Lett.* **4**, 1150–1157 (2019).
85. W. Li, *et al.*, Polarons in perovskite solar cells: effects on photovoltaic performance and stability. *J. Phys. Energy* **5**, 024002 (2023).
86. M. J. Schilcher, *et al.*, The Significance of Polarons and Dynamic Disorder in Halide Perovskites. *ACS Energy Lett.* **6**, 2162–2173 (2021).
87. Z. Guo, J. Wang, W. J. Yin, Atomistic origin of lattice softness and its impact on structural and carrier dynamics in three dimensional perovskites. *Energy Environ. Sci.* **15**, 660–671 (2022).
88. C. M. Mauck, W. A. Tisdale, Excitons in 2D Organic–Inorganic Halide Perovskites. *Trends Chem.* **1**, 380–393 (2019).
89. D. A. Egger, *et al.*, What Remains Unexplained about the Properties of Halide Perovskites? *Adv. Mater.* **30**, 1800691 (2018).

90. Z. Chen, Y. Guo, E. Wertz, J. Shi, Merits and Challenges of Ruddlesden–Popper Soft Halide Perovskites in Electro-Optics and Optoelectronics. *Adv. Mater.* **31**, 1803514 (2019).
91. N. J. Weadock, *et al.*, The nature of dynamic local order in $CH_3NH_3PbI_3$ and $CH_3NH_3PbBr_3$. *Joule* **7,** 1051-1066 (2023).
92. K. T. Munson, E. R. Kennehan, G. S. Doucette, J. B. Asbury, Dynamic Disorder Dominates Delocalization, Transport, and Recombination in Halide Perovskites. *Chem* **4**, 2826–2843 (2018).

# Supporting Information

# Experimental Evidence for Defect Tolerance in Pb-Halide Perovskite

Naga Prathibha Jasti[1,2], Igal Levine[3], Yishay (Isai) Feldman[2], Gary Hodes[2], Sigalit Aharon[2]* and David Cahen[1,2]*

[1]*Bar Ilan University, Ramat Gan, 5290002, Israel*
[2]*Weizmann Institute of Science, Rehovot, 7610001, Israel*
[3]*Division Solar Energy, Helmholtz-Zentrum Berlin für Materialien und Energie GmbH, Berlin, 12489, Germany*

## S1. MATERIALS PREPARATION PROCEDURES

**Synthesis of MAPbI$_3$ Single Crystals:**

The high-quality MAPbI$_3$ (MAPI) single crystals were prepared by inverse temperature crystallization (1). The higher-quality MAPI (MAPI-HQ) crystals were prepared from 1:1 of MAI:PbI$_2$ (>99.99 %, Greatcell Solar Materials, 99 %, Merck, respectively), 1.4 M solution in gamma-butyrolactone (99+%, Alfa Aesar), under ambient atmosphere. The precursor solution was stirred at 75°C, followed by filtration using PTFE (hydrophilic) filters (pore size: 0.22 µm) into a crystallization dish with a seed crystal (∼1-2 mm in size). The single crystals (∼0.7-1cm in size) formation occurred by ramping the temperature of the oven from 75°C to 120°C in 18 hours. The crystals were extracted by cleaning with filter paper placed on a hot plate at 110°C.

The lower-quality MAPI (MAPI-LQ) crystals were prepared from 1.1:1 of MAI:PbI$_2$ (>99.99 %, Greatcell Solar Materials, 99 %, Merck, respectively), 1.1: 1 M solution, respectively, in gamma-butyrolactone (99+%, Alfa Aesar), under ambient atmosphere. The precursor solution was stirred at 60°C, followed by filtration using PTFE (Hydrophilic) filters (pore size: 0.450 µm) into an open vessel to crystallize at 60°C. The solution temperature was increased to 90°C over the course of 1 hour and then to 110°C, over 3 hrs to promote the evaporation of the solvent. The crystallization was stopped just before the level of the solution reached the crystal surface so as to avoid exposure of the crystal surface to air. The crystals were then harvested from the solution and, without rinsing, immediately dried with absorbent paper. Crystals with a size of 0.5-1 cm were obtained.

**Synthesis of BA$_2$PbI$_4$ (C4N1), BA$_2$MAPb$_2$I$_7$ (C4N2) and BA$_2$MA$_2$Pb$_3$I$_{10}$ (C4N3) Single Crystals:**

Butylammonium lead iodide (BA$_2$PbI$_4$, C4N1), butylammonium methylammonium lead iodide ((BA$_2$MAPb$_2$I$_7$, C4N2), (BA$_2$MA$_2$Pb$_3$I$_{10}$, C4N3)) were crystallized using the slow cooling method (2). 5.045 mmol (1.126 g) PbO (ACS reagent, 99.0 %, Sigma-Aldrich) was dissolved in 5mL HI (57 % in H$_2$O, Sigma Aldrich, stored at 4°C) and 850 µL hypophosphorous acid solution (50 wt.% in H$_2$O, Sigma-Aldrich) in a 20 mL vial by heating and stirring on a hot plate that was set to 110°C. The color of the mixture changed from black to clear yellow within a minute. The stirring and heating continued until the full dissolution of the PbO (1-2 hours). In the meantime, in an ice-bath, 3 mL of HI (4°C) were mixed with 494 μL or 98 μl butylamine (99.5 %, Sigma-Aldrich) for preparation of C4N1 and C4N3, respectively, by vigorously stirring the HI with a magnetic stirrer and adding the butylamine dropwise. In the case of C4N3 preparation, 0.477 g methylammonium iodide (MAI, GreatCellSolar materials), was added to the HI prior to the butylamine addition. The vial was then tightly sealed and stirring continued until no vapor was seen in the upper part of the vial. Once the two above mixtures were ready, the HI+butylamine (+MAI, for C4N2 and C4N3) mixture was added, dropwise, to the Pb-containing vial while continuing to stir and heat. This led to the formation of an orange powder in the vial of C4N1 and a dark red powder in the vial of C4N2 and C4N3. Then the vial was tightly sealed and heated, while stirring, until the orange powder fully dissolved (the hotplate was set for this purpose to 140-170°C for 10-20 minutes). Once the solution was perfectly clear, the magnetic stirrer was carefully removed, the vial tightly sealed again, and transferred into an oven for controlled slow cooling, which was preheated to 105°C. The temperature of the oven was gradually decreased to RT at a rate of 1°C /h. Once the cooling process was completed, large orange plates of C4N1 crystals and reddish-black plates of C4N2 and C4N3 crystals were seen in the bottom of the vial. C4N2 and C4N3 crystals were both obtained from the same solution. The single crystals were then taken out by evacuating the supernatant and drying them gently with filter paper. To further remove any traces of the solvents, the crystals were then dried in the vacuum oven at 40°C for 12 hours.

## S2. CHARACTERIZATION PROCEDURES

**X-Ray diffraction characterization for powder and single crystals:**

XRD characterization was carried out in reflection geometry using Rigaku (Tokyo, Japan) θ-θ diffractometers: an Ultima III (for powdered samples, prepared by crushing single crystals; cf. Fig. S2) equipped with a sealed Cu anode tube operating at 40 kV/40 mA and a TTRAX III (for single crystals) equipped with a rotating Cu anode X-ray tube operating at 50 kV/200 mA. The scintillation detector was aligned to intersect the diffracted beam after it passed a bent graphite monochromator to remove Kβ radiation. For powder samples (pulverized single crystals), a θ-2θ scan was performed in the 2θ range of 2°- 80° with a step size of 0.025° and a scan rate of 1° per minute.

For single crystals, first, using open slits after a sample, a θ-2θ scan was performed in the 2θ range of 5°- 60° (for MAPbI$_3$) and 1.5° to 40° (for C4N1, C4N2, C4N3) with a step size of 0.01° and a scan rate of 10° per minute. Then the rocking curves (illustration in Fig. S5) were measured at angles of 2θ = 40.49° (for MAPbI$_3$), 2θ = 25.75° (for C4N1), 2θ = 27.15° (for C4N2) and 2θ = 27.52° (for C4N3) in the θ-2θ scan and corresponding to the Bragg condition for diffraction from the (400) and (220) (for MAPbI$_3$), (008) (for C4N1), (1200) (for C4N2) and

# Supporting Information

(0160) (for C4N3) crystallographic planes, parallel to the sample surface. Keeping 2θ constant, the source and the detector stage were synchronously moved (ω/θ scan) along the arc, while the sample remained in a fixed position for each ϕ position. To reduce the RC broadening due to the divergence of the primary beam, an incident slit 0.1 mm wide was used. As a comparison using the same measurement conditions, the full width at half maximum (FWHM) of single crystal silicon (111) was 0.027°. XRD analysis was performed using the Jade Pro (MDI, US) and Origin software packages. NOTE: the asymmetric tails in the diffractograms, shown in Fig. S2a,b,c for the low angle peaks, are the result of axial divergence due to the relatively wide (5°) Soller slits (3) we needed on this (Ultima) instrument to get good signal to noise.

**Photoluminescence (PL) measurements:**
The single crystal samples were excited at 457 nm wavelength using a diode-pumped laser (Cobolt DPL-08). An ultra-steep long-pass edge filter Semrock - 458 NM Razoredge (REL 458 nm) was used to filter the excitation laser wavelength. A 10X objective (NA = 0.3) with a collimator was used for exciting and collecting the PL signal. The PL emission spectra were collected using a CCD-based spectrometer (Ocean Optics S2000).

**Time-resolved PL measurements:**
The PL-lifetime measurements that are reported, were conducted with a Horiba Scientific Fluoromax-4 spectrofluorometer. The sample was excited by a 375 nm Nano-LED, with ~2.5 x $10^9$ photons/cm$^2$ flux, 45 ps pulse duration, and 1 MHz repetition rate (peak power of 300 mW). The emission was collected at three different wavelengths, 520 nm for C4N1, 615 nm for C4N3, and 780 nm for the MAPI samples, using Time-Correlated Single-Photon Counting (TCSPC) detection.

Results similar to the ones obtained with the Horiba system, were obtained with a home-built system, using a ps pulsed diode laser (Picoquant LDH-P-C-640B). The photon flux of the pulse was 5 x $10^{10}$ photons/cm$^2$, the pulse duration 2-5 ns and the repetition rate 1 MHz. The PL emission spectra and the PL decay characteristics were recorded using a photomultiplier tube (Hamamatsu model H10721-01). For each of the HaP compositions, the TRPL data were collected by using a monochromator (Shamrock-Andor, SR-193i-Bi). For these measurements the C4N1 and C4N3 crystals were excited at 450 nm, and the MAPI crystals at 640 nm.

**Transmission optical microscope imaging:**
The microscope images were collected using Olympus BX51 light microscope, equipped with an Olympus U-AN360 rotatable analyzer slider.

**Modulated Surface Photovoltage (SPV) measurements:**
Modulated SPV spectra were measured in the configuration of a parallel plate capacitor (quartz cylinder partially coated with a $SnO_2$:F electrode, cover glass #0 as an insulator), as described in references (4, 5) under ambient conditions. Layered crystals were first peeled off

# Supporting Information

using a 3M scotch tape prior to the SPV measurements. The SPV signal is defined as the change in the surface potential as a result of the illumination. In our case, the illumination was provided by a halogen lamp, coupled to a quartz prism monochromator (SPM2), and modulated at a frequency of 8 Hz by using an optical chopper. In order to avoid artifacts due to stray light in the sub-bandgap region, an appropriate long pass filter was used, depending on the bandgap of the sample (6). In-phase and 90° phase-shifted SPV signals were detected with a high-impedance buffer and a double-phase lock-in amplifier (EG&G 5210). The in- and out of- phase SPV signals correspond to the fast and slow SPV responses, relative to the light modulation period, respectively. A positive sign of the in-phase signals means that photo-generated electron-hole pairs are separated with preferentially electrons towards the bottom substrate, while a negative sign of the in-phase signal means that the photo-generated electron-hole pairs are separated with preferentially electrons towards the top electrode. The amplitude of the modulated SPV signal is defined as the square root of the sum of the squared in-phase and 90° phase-shifted SPV signals (5).

## S3. SUPPLEMENTARY FIGURES AND DATA

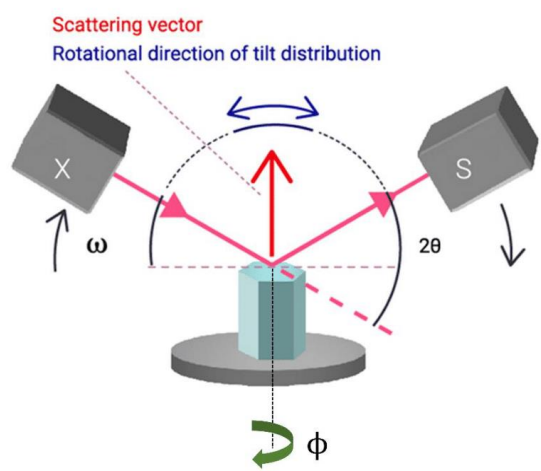

**FIG. S1.** Schematic diagram illustrating the rocking curve measurement (ω-scan) and the φ direction.
(Adapted from https://www.rigaku.com/applications/bytes/xrd/smartlab/57445041)

In an ω-scan, the sample is rotated around the ω-axis, and the plot of the scattered X-ray intensity as a function of ω is recorded; the resulting plot is called a 'rocking curve' (RC). Practically, this is done by keeping the sample fixed while scanning the X-ray source, X, and detector, S, concurrently toward the same direction, such that the incident angle ω is varied but 2θ remains unchanged. The FWHM and the tails of the RCs contain information on the 1D & 2D, and 0D, 1D defects. RC profile comparisons, along with use of reciprocal space mapping to measure diffuse X-ray scattering can help detect differences in point defect types and densities (cf. (7) and references therein).

# Supporting Information

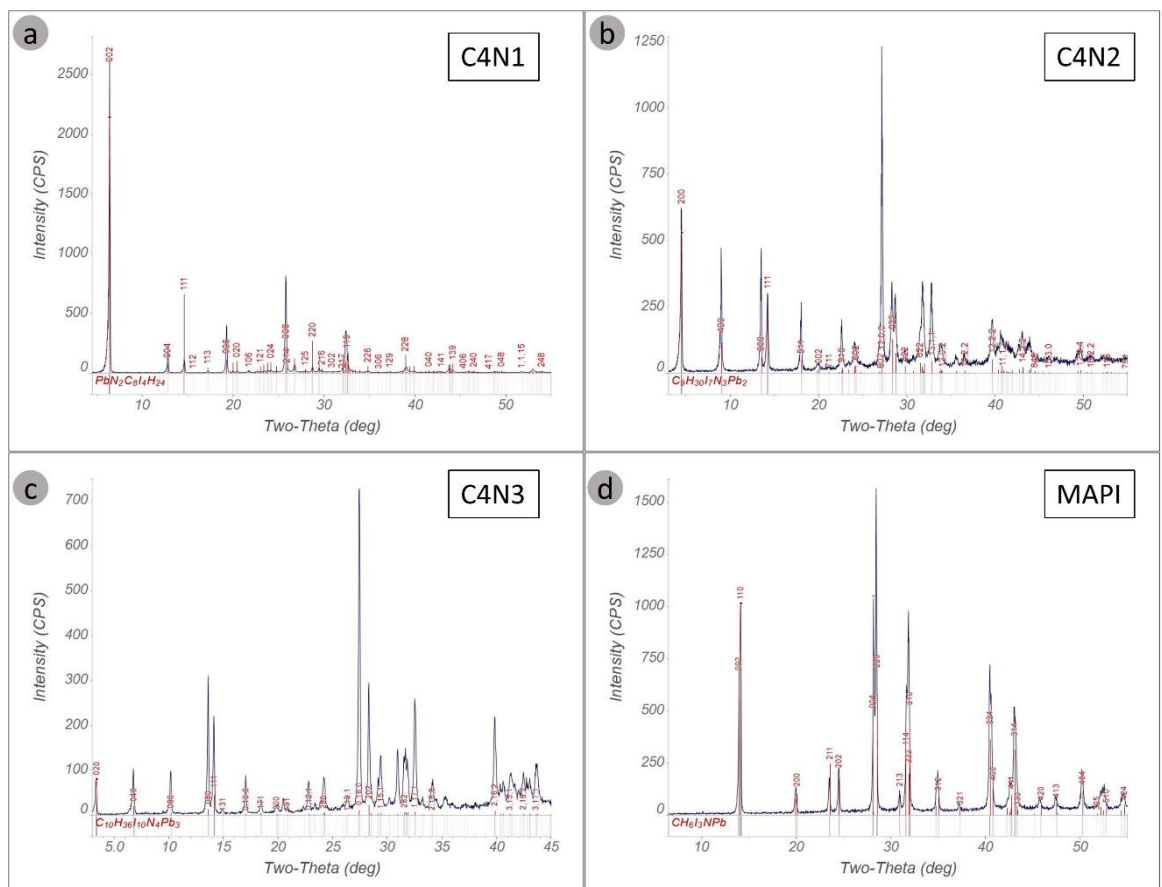

**FIG. S2**. Powder XRD pattern of a pulverized single crystal of a. $BA_2PbI_4$ (C4N1); b. $BA_2MAPb_2I_7$ (C4N2); c. $BA_2MA_2Pb_3I_{10}$ (C4N3); and d. $MAPbI_3$ (MAPI).

Samples used for powder XRD were prepared by pulverising the single crystals. This is needed to check the phase purity of the single crystals that are used in this study. The powder XRD patterns, presented in Figs. S2 a-d show that all samples (C4N1, C4N2, C4N3 and MAPI) are phase pure, i.e., without detectable secondary phases. Moreover, with the exception of MAPI, all XRD patterns show a preferential orientation (texture) of some crystallographic orientation (C4N1 – *(00ℓ)*, C4N2 – (*h00*) and C4N3 – *(0k0)*) of the measured powders (the corresponding experimental peaks are much higher than their database values, red lines). This is most likely due to the fact that during the preparation of the sample these particular planes lay parallel to the surface, due to the anisotropic morphology of the particles.

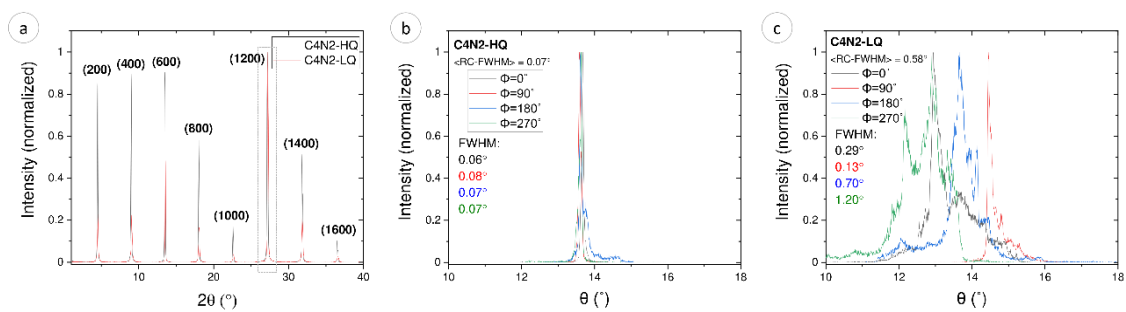

**FIG. S3.** a. θ-2θ XRD pattern for $BA_2MAPb_2I_7$ (C4N2), HQ and LQ. The peaks marked with the dashed rectangle is the diffraction angle (27.15°) at which the rocking curve, RC, was recorded; b. Rocking curves (ω scan) of a C4N2-HQ crystal performed at different φ values; c. Rocking curves of a C4N2-LQ crystal performed at different φ values.

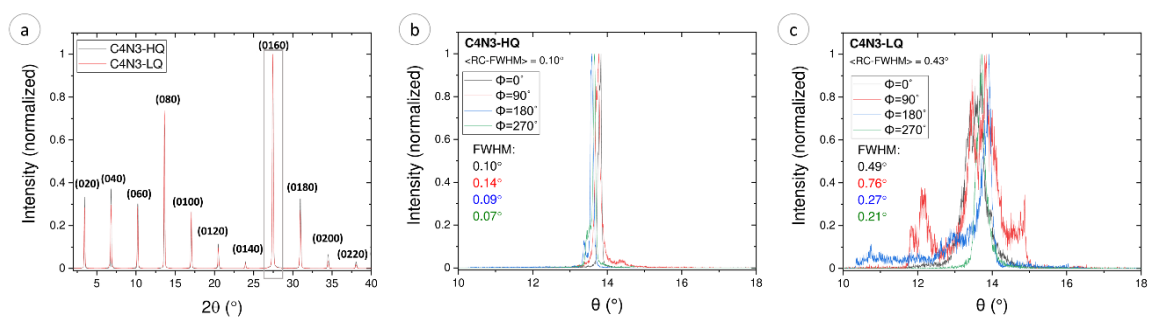

**FIG. S4**. a. θ-2θ XRD pattern for $BA_2MA_2Pb_3I_{10}$ (C4N3), HQ and LQ. The peaks marked with the solid rectangle is the diffraction angle (27.52°) at which the rocking curve, RC, is recorded; b. Rocking curves (ω scan) of a C4N3-HQ crystal performed at different φ values; c. Rocking curves of a C4N3-LQ crystal performed at different φ values.

For C4N2-HQ (Figure S3b, SI) and C4N3-HQ (Figure S4b, SI) in the ω-scan (RC), and at different φ angles, the average FWHM value was relatively small (0.07°- 0.1°), but for C4N2-LQ (Figure S3c, SI) and C4N3-LQ (Figure S4c, SI), the average FWHM values were higher (0.4° – 0.6°), and multiple peaks were detected. These results indicate that the LQ crystals have a significant structural disorder and relatively poor crystal quality.

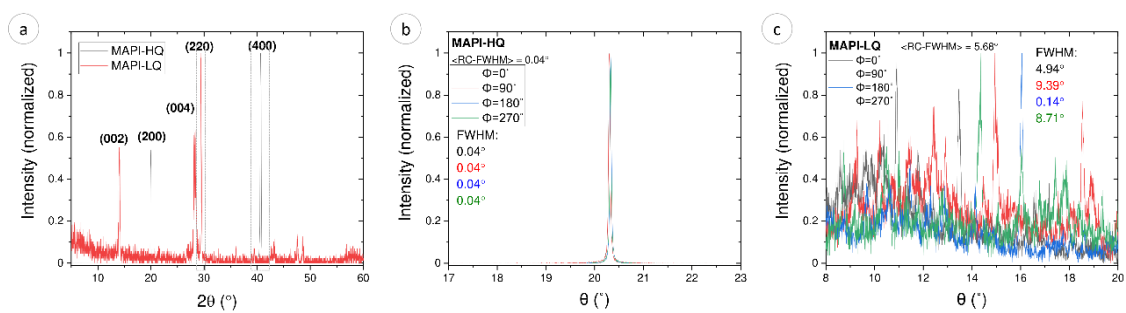

**FIG. S5**. a. θ-2θ XRD patterns for $MAPbI_3$ (MAPI), HQ and LQ. The peaks marked with the dashed rectangles are the diffraction angles (29.44° for MAPI-LQ, and 40.60° for MAPI-HQ) at which the rocking curve, RC, was recorded; b. Rocking curves (ω-scan) of a MAPI-HQ crystal performed at different φ values; c. Rocking curves of a MAPI-LQ crystal performed at different φ values.

In the case of 3D MAPI crystals, for MAPI-HQ (Figure S5b, SI), the ω-scan (RC) at different φ angles showed that the average FWHM values were very small (0.04°), while for MAPI-LQ (Figure S5c, SI) the average FWHM values were very large (∼5.68°) and indicate that the LQ crystals have a significant structural disorder and relatively poor crystal quality.

**Supporting Information**

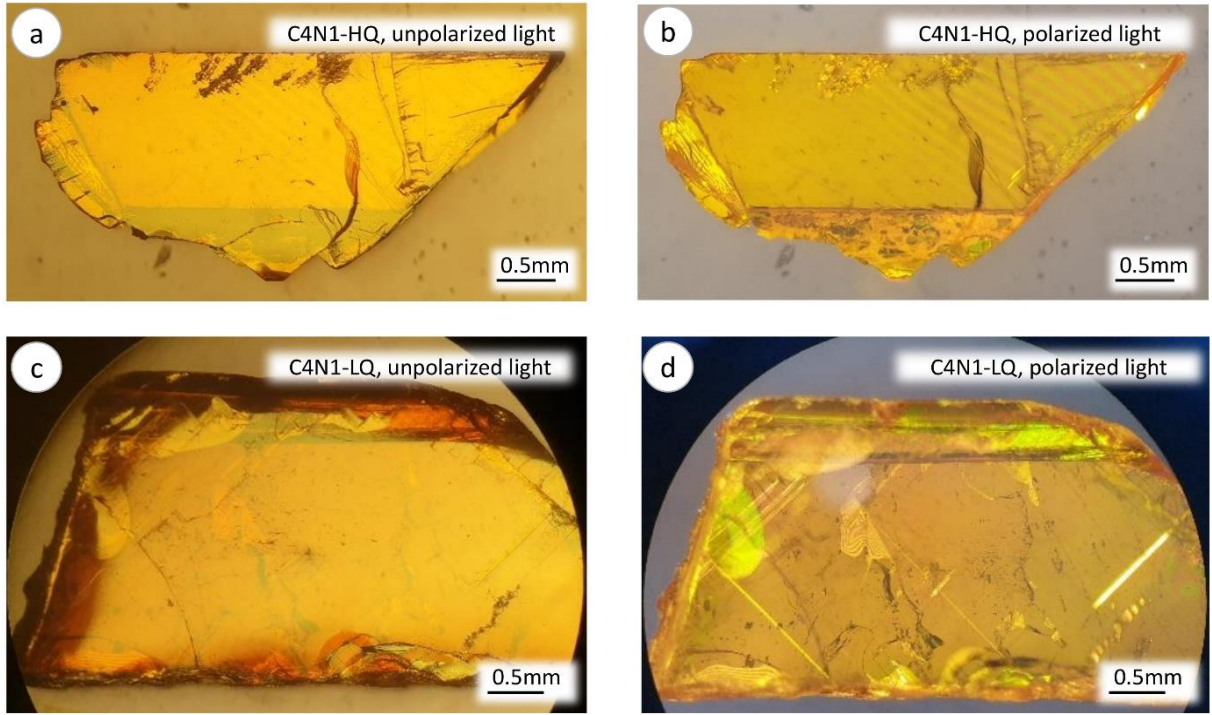

**FIG. S6**. Transmission optical microscope images (magnified view) of a. C4N1-HQ under unpolarized light, b. C4N1-HQ under polarized light, c. C4N1-LQ under unpolarized light, and d. C4N1-LQ under polarized light.

From Figures S6 a and c, under unpolarized light, we see that with low magnification, the differences in crystal quality between HQ and LQ crystals are not clearly visualized.

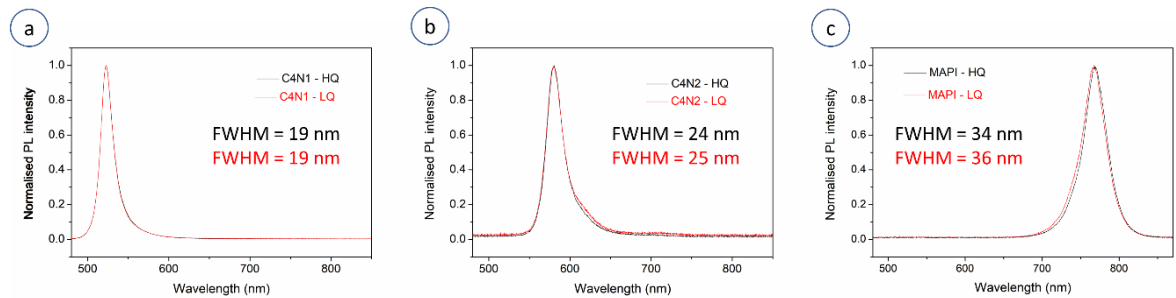

**FIG. S7.** PL spectra for HQ and LQ crystals a. C4N1 b. C4N2 c. MAPI

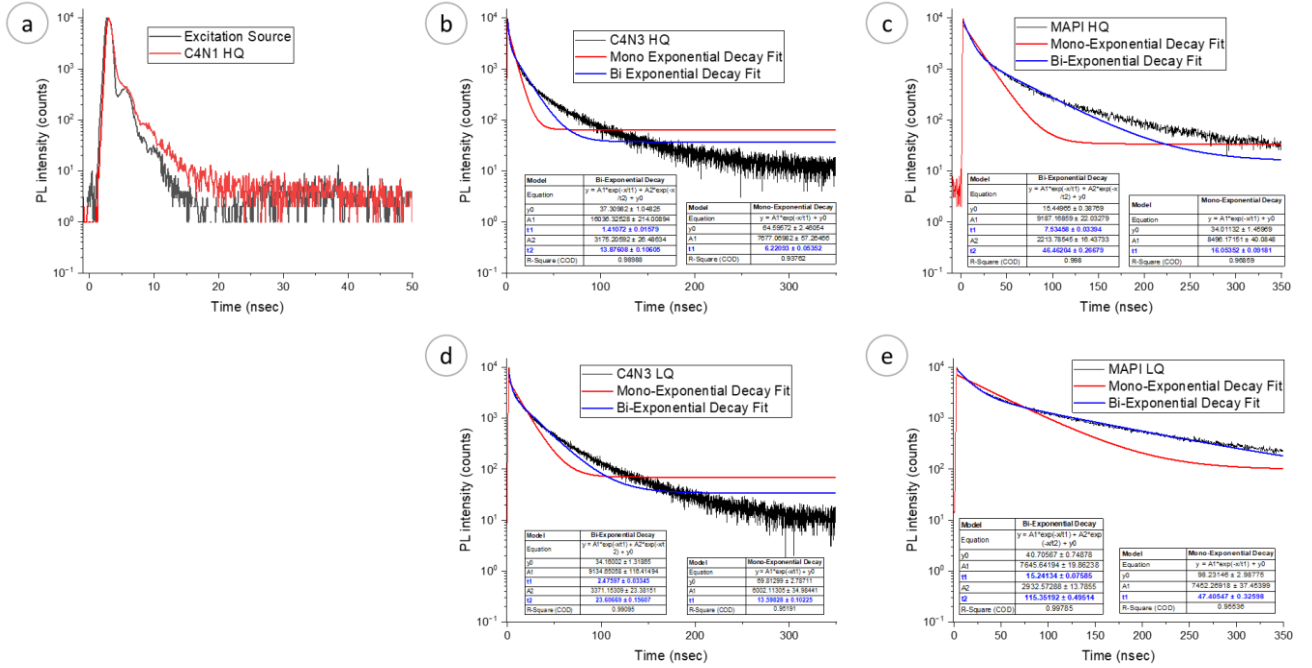

**FIG. S8.** Time-resolved photoluminescence (TRPL) traces and the corresponding decay times of photoexcited carriers (blue font in the table), extracted from attempts at mono- and bi-exponential fits to the traces: (a) the excitation source *vs.* the signal of C4N1-HQ (b) C4N3-HQ; (c) MAPI-HQ; (d) C4N3-LQ; (e) MAPI-LQ.

The decay curves in figure S8 above clearly show that:
1- The C4N1 signal nearly overlaps with the excitation pulse, which prevents extracting f a meaningful lifetime.
2- Both the mono- and bi-exponential decay functions, used to fit the data from the HQ and LQ C4N3 and MAPI single crystals, result in poor fits to the data. This result means that the lifetimes extracted from them are not entirely reliable; this deviation points to a higher complexity of the TRPL data analysis here.

*Table S1: The lifetimes that were extracted from the PL decay curves of the different crystals, by either mono-exponential decay (Excitation source and C4N1) or bi-exponential decay (C4N3 and MAPI).*

| Sample | $\tau_1$ [nsec] | $\tau_2$ [nsec] |
|---|---|---|
| Excitation source | $0.630 \pm 0.005$ | - |
| C4N1 HQ | $0.565 \pm 0.002$ | - |
| C4N3 HQ | $1.41 \pm 0.02$ | $13.9 \pm 0.1$ |
| C4N3 LQ | $2.48 \pm 0.03$ | $23.6 \pm 0.2$ |
| MAPI HQ | $7.54 \pm 0.03$ | $46.5 \pm 0.3$ |
| MAPI LQ | $15.24 \pm 0.08$ | $115.4 \pm 0.5$ |

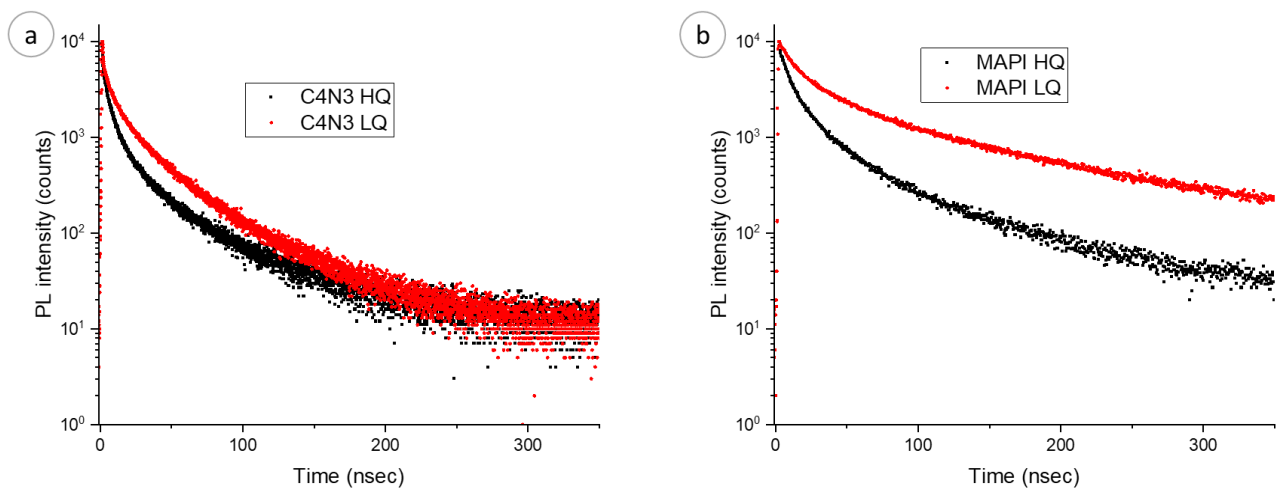

**FIG. S9.** Time-resolved photoluminescence (TRPL) traces of photoexcited carriers (at maximum PL emission wavelengths, for C4N3 and for MAPI, both HQ and LQ crystals).

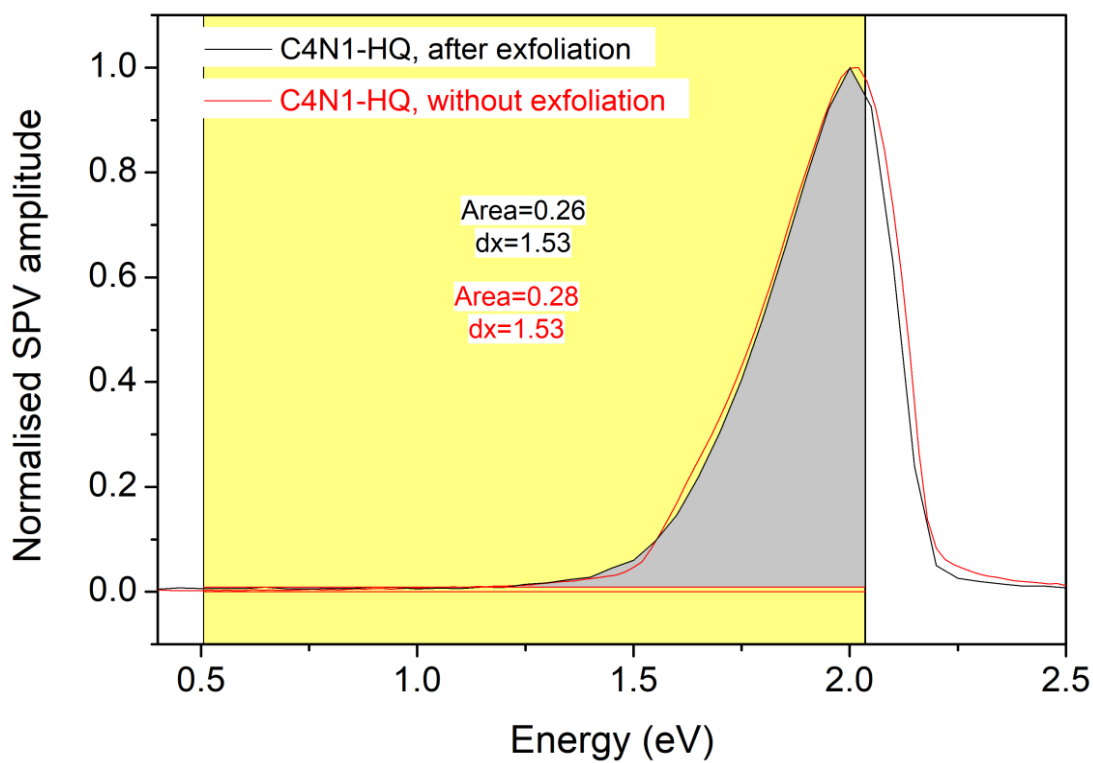

**FIG. S10**. Modulated sub-band-gap SPV spectra for C4N1 HQ crystal with and without surface exfoliation before the measurement. Area represents the area under the curve (grey zone), calculated between the limits of the rectangle edges, (dx) (yellow zone), with the baseline (lower limit on y scale) set at the SPV noise level. For clarity only one such grey zone corresponding to black spectrum is shown here. On the x (= photon energy) axis the upper limit corresponds to the cut-on energy of the 610nm long-pass filter used for C4N1 and the

# Supporting Information

lower limit is around 0.5 eV, the photon energy, where SPV signal disappears in the noise level. In the legend dx is the difference between the low and high photon energy limits, used to calculate the Area. See section 4 for further error calculation details.

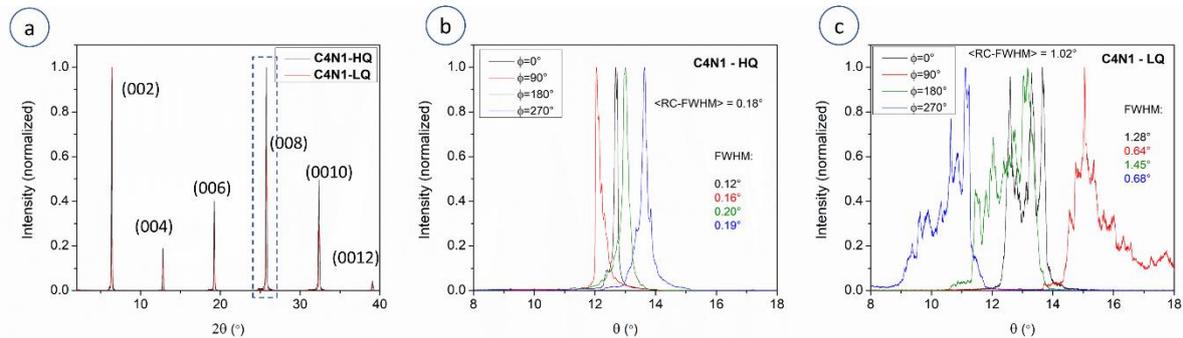

**FIG. S11**. a. θ-2θ XRD patterns for $BA_2PbI_4$ (C4N1), HQ and LQ crystals, used for SPV measurements, discussed in Table 1, main text. The peaks marked with the dashed rectangle are the diffraction angle (27.45°) at which the rocking curve, RC, was recorded; b. Rocking curves (ω-scan) of a C4N1-HQ crystal performed at different ϕ values. c. Rocking curves of a C4N1-LQ crystal at different ϕ values.

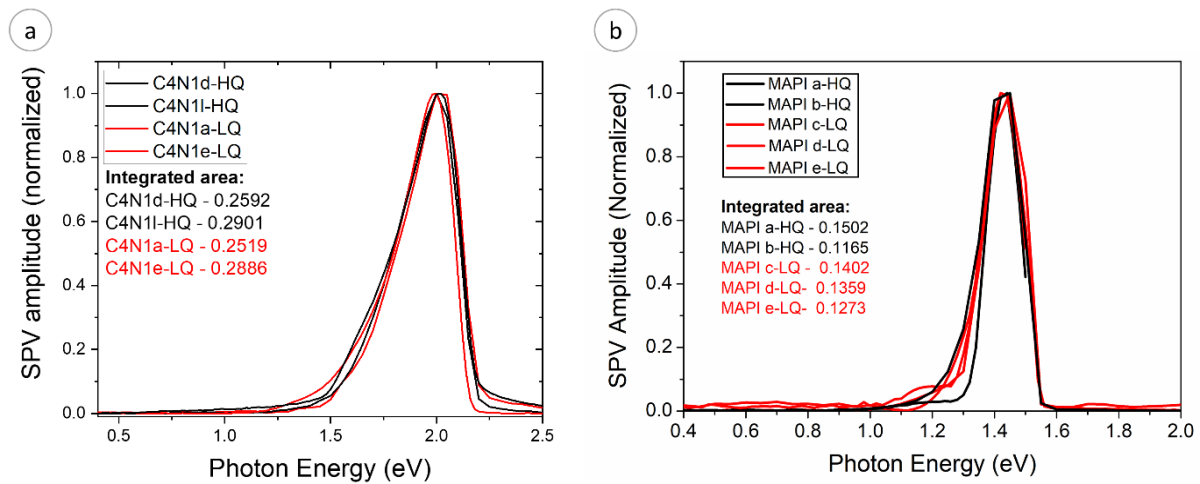

**FIG. S12**. The normalized SPV amplitude vs. photon energy for HQ (black traces) and LQ (red traces) single crystals of (a) C4N1 and (b) MAPI.

## S4. CALCULATION DETAILS FOR INTEGRATED AREA OF THE SUB-BANDGAP SPV SPECTRUM AND THE CORRESPONDING ERROR:

To avoid artifacts due to stray light in the sub-bandgap region, an appropriate long pass filter was used (610 nm for C4N1, 750 nm for C4N2, 750 nm for C4N3 and 850 nm for MAPI), depending on the bandgap of the sample (C4N1: 2.4 eV, C4N2: 2.1 eV, C4N3: 2.0 eV and MAPI: 1.6 eV). For each sample, integrating the area under the curve of the sub-bandgap SPV spectrum was done between two limits that were set as follows:

# Supporting Information

1. The energy corresponding to the long pass filter is taken as the upper limit of the energy, and

2. The energy where the SPV signal drops so that the noise level starts to dominate is set as the lower limit of the energy.

The energy error of the limits translates into an error for the integrated area; it is estimated from the equation , by considering the error in wavelength of the monochromator ($d\lambda$ = 5 nm). To estimate the error in the SPV values, we found that the noise levels amounted to a maximum of 0.01% of the peak SPV signals. Using these values, the error in the integrated area is estimated to be, at the very most 4% for the results given in Table 1.

**S5. VIDEOS THAT DEMONSTRATE THE QUALITATIVE STRUCTURAL EVALUATION OF 3D MAPI CRYSTALS ARE REPORTED IN SUPPLEMENTARY FILES:**

1. Supporting Video 1 demonstrates the mechanical cleaving of a MAPI-HQ crystal, which leads to the exposure of mirror-like crystal faces of the two freshly exposed surfaces.

2. Supporting Video 2 demonstrates the mechanical cleaving of a MAPI-LQ crystal, which leads to multiple fragments of the crystal.

**References:**

1. M. I. Saidaminov, A. L. Abdelhady, G. Maculan, O. M. Bakr, Retrograde solubility of formamidinium and methylammonium lead halide perovskites enabling rapid single crystal growth. *Chem. Commun.* **51**, 17658–17661 (2015).
2. S. Aharon, *et al.*, 2D Pb-Halide Perovskites Can Self-Heal Photodamage Better than 3D Ones. *Adv. Funct. Mater.* **32**, 2113354 (2022).
3. H. Klug, L. Alexander, X-ray Diffraction Procedures: For Polycrystalline and Amorphous Materials, 2nd Edition. *Willey, New York, EUA*, 992 (1974).
4. T. Dittrich, *et al.*, Comparative study of $Cu(In,Ga)Se_2$/CdS and $Cu(In,Ga)Se_2$/$In_2S_3$ systems by surface photovoltage techniques. *Thin Solid Films* **535**, 357–361 (2013).
5. T. Dittrich, S. Fengler, Surface Photovoltage Analysis of Photoactive Materials. *Surf. Photovoltage Anal. Photoactive Mater.* (2020).
6. I. Levine, *et al.*, Deep Defect States in Wide-Band-Gap $ABX_3$ Halide Perovskites. *ACS Energy Lett.* **4**, 1150–1157 (2019).
7. S. Dolabella, *et al.*, Lattice Strain and Defects Analysis in Nanostructured Semiconductor Materials and Devices by High-Resolution X-Ray Diffraction: Theoretical and Practical Aspects. *Small Methods* **6**, 2100932 (2022).